\documentclass[12pt]{iopart}

\usepackage{graphicx}
\usepackage{url}
\usepackage[usenames,dvipsnames]{xcolor}
\usepackage[normalem]{ulem}
\usepackage{CJKutf8}

\newcommand{\cntext}[1]{\begin{CJK*}{UTF8}{bsmi}#1\end{CJK*}}

\begin{document}

\title[A Comparison of transport methods for CCSNe]
{The Impact of Different Neutrino Transport Methods on Multidimensional Core-collapse Supernova Simulations}

\author{Kuo-Chuan Pan (\cntext{潘國全})$^{1,2,3}$, Carlos Mattes$^{4}$, Evan P. O'Connor$^{5}$, Sean
M. Couch$^{1,2,6,7}$, Albino Perego$^{8,9}$, Almudena Arcones$^{4,10}$}

\address{
  $^{1}$ Department of Physics and Astronomy, Michigan State University, East Lansing, MI 48824, USA\\
  $^{2}$ Joint Institute for Nuclear Astrophysics-Center for the Evolution of the Elements, Michigan State University, East Lansing, MI 48824, USA\\
  $^{3}$ Department of Physics and Institute of Astronomy, National Tsing Hua University, Hsinchu, Taiwan\\
  $^{4}$ Institut f\"ur Kernphysik, Technische Universit\"{a}t Darmstadt, Schlossgartenstrasse 9, D-64289 Darmstadt, Germany\\
  $^{5}$ Department of Astronomy and Oskar Klein Centre, Stockholm University, AlbaNova, SE-106 91 Stockholm, Sweden \\
  $^{6}$ Department of Computational Mathematics, Science, and Engineering, Michigan State University, East Lansing, MI 48824, USA\\
  $^{7}$ National Superconducting Cyclotron Laboratory, Michigan State University, East Lansing, MI 48824, USA\\
  $^{8}$ Istituto Nazionale di Fisica Nucleare, Sezione Milano Bicocca, gruppo collegato di Parma, I-43124 Parma, Italia\\
  $^{9}$ Universit\`a degli Studi di Milano-Bicocca, Dipartimento di Fisica, Piazza della Scienza 3, 20126 Milano, Italia.\\
  $^{10}$ GSI Helmholtzzentrum f\"ur Schwerionenforschung GmbH, Planckstr. 1, Darmstadt 64291, Germany\\
}
\ead{pankuoch@msu.edu}
\vspace{10pt}
\begin{indented}
\item[]April 2018
\end{indented}
\begin{abstract}

Neutrinos play a crucial role in the core-collapse supernova (CCSN)
explosion mechanism.
The requirement of accurately calculating the transport of neutrinos makes simulations of the CCSN mechanism extremely challenging and computationally expensive.
Historically, this stiff challenge has been met by making approximations to the full transport equation.
In this work, we compare CCSN simulations in one- and two-dimensions
with three approximate neutrino transport schemes, each implemented in the {\tt FLASH} simulation framework.
We compare a two moment M1 scheme with an analytic closure (M1),
the Isotropic Diffusion Source Approximation (IDSA),
and the Advanced Spectral Leakage (ASL) method.
We identify and discuss the advantages and disadvantages of each scheme.
For each approximate transport scheme, we use identical grid setups, hydrodynamics, and gravity solvers
to investigate the transport effects on supernova shock dynamics
and neutrino quantities.
We find that the transport scheme has a small effect
on the evolution of protoneutron star (PNS) radius, PNS
mass, and the mass accretion rate.
The neutrino luminosities, mean energies, and shock radii have a $\sim$
10-20\% quantitative difference but the overall qualitative trends
are fairly consistent between all three approximations.
We find larger differences in the gain region properties, including the
gain region mass and the net heating rate in the gain region, as well
as the strength of PNS convection in the core.
We investigate the progenitor, nuclear equation of state, and
stochastic perturbation dependence of our simulations and find
similar magnitudes of impact on key quantities.
We also compare the computational expense of the various approximations.

\end{abstract}

%
%
\section{Introduction}

Simulations of the core-collapse supernova (CCSN) mechanism require a challenging array of input physics, including multidimensional magnetohydrodynamics, detailed neutrino transport, involved microphysics, and general relativistic gravity.
Over the roughly six-decade history of computational investigation of the CCSN mechanism, this complex mix of input physics has put this problem at the cutting-edge of computational complexity and expense.
Dominating the difficulty is the requirement to accurately simulate the transport of neutrinos in this complex context.
The transport of neutrinos must be followed from the diffusive, optically thick protoneutron star (PNS), through the semi-transparent region between the PNS and the stalled supernova shock, and into the completely transparent, free-streaming region beyond the shock.
Solving the full seven-dimensional Boltzmann transport equation for
neutrinos is almost universally too steep a challenge for
high-resolution, time-dependent CCSN simulations (but see \cite{nagakura:2017}).
Approximating the full transport equation is common, though different groups working on the problem apply various different approaches.

Owing to the challenge of CCSN mechanism simulations, tremendous progress has been made in recent years.
After decades of research marked by cycles of promise then failure, several groups are now reporting successful neutrino-driven explosions in 2D \cite{Marek:2009, Muller:2012a, Bruenn:2016,Burrows:2016,Summa:2016,Radice:2017,oconnor:2018,Skinner:2016,vartanyan:2018}.
These results often show significant quantitative, and even qualitative, differences for similar initial conditions.
These various works use a variety of different hydrodynamic and neutrino transport approaches, making a direct code-to-code verification impossible.
Here, we present a controlled code-to-code verification of different neutrino transport approximations commonly used in multidimensional simulations of the CCSN mechanism.
We compare the two-moment explicit ``M1'' closure method \cite{OConnor:2015}, the isotropic diffusion source approximation (IDSA) \cite{Liebendorfer:2009}, and the advanced spectral leakage (ASL) method \cite{Perego:2014}, using the same simulation framework, FLASH \cite{fryxell:2000, dubey:2009}.
Among these three transport methods, the M1 scheme, which solves the first two moments of the Boltzmann equation, is the most accurate but also the most computationally expensive. 
The IDSA scheme assumes that the transported particles can be decomposed into trapped and free streaming components and solved separately. The trapped component is similar to a flux limited diffusion approach (or M0) but the flux factor in the streaming region is improved by the flux from the streaming component.   
The ASL scheme, which is based on a multi-energy leakage method, including spectral information about trapped neutrinos, is the most approximate and efficient scheme. 
The detailed description of each approach will be presented in Section~\ref{sec:transport}.

In this study, we restrict ourselves to 1D and 2D simulations.
While current high-fidelity simulations in 1D only result in explosions for very low-mass iron core progenitors \cite{kitaura:2006}, it has long been understood that multidimensional effects in 2D, such as neutrino-driven convection and the standing accretion shock instability (SASI), aid shock expansion and explosion \cite{burrows:1995}.
In recent years, it has become clear that fully 3D simulations are necessary as the enforced symmetry of 2D is unduly influencing the quantitative outcomes of CCSN simulations \cite{couch:2013a, hanke:2013, lentz:2015, melson:2015a, melson:2015, Kuroda:2016, Kuroda:2017, Ott:2018, OConnor:2018c}.

While a controlled comparison in 3D is desirable, there is yet no
detailed code comparison of CCSN simulation in even 2D (although see the recent work of \cite{Just:2018,Cabezon:2018}).
The careful comparison of 1D, high-fidelity CCSN simulation codes executed by \cite{liebendorfer:2005a} is extremely valuable, and even led to improvements in the approach for approximating general relativistic gravity now widely used \cite{marek:2006}.
Multidimensional comparisons are challenging because of the non-linear feedback between the hydrodynamics, transport, and gravity.
The development of non-linear instabilities in 2D and 3D can make it a challenge to disentangle what differences in the underlying numerical methods are leading to differences in the results. 
Here, we attempt to address this difficulty by carrying out 1D and 2D comparisons of different neutrino transport approximations using {\it identical} hydrodynamics, gravity solvers, EoS implementations, and computational grids.
Non-linear feedback and instabilities can still magnify small differences, but using a common code for everything except neutrino transport allows us the most controlled study of the impacts of different transport methods on the overall results of CCSN simulations.
This approach can shed light on the magnitude of the impact of different transport approximations in multidimensional simulations and their relative computational expense.

Our presentation of this study is as follows.
In Section \ref{sec:methods}, we describe our numerical methods, including the details of the three different transport approximations we employ.
In Section \ref{sec:results}, we present our results, starting with a comparison of 1D simulations then continuing on to discuss 2D simulations.
We also briefly explore the impact of different equations of state (EoS) and contrast this with the differences arising from different transport schemes.
Additionally, in Section \ref{sec:results}, we discuss the difference in overall code performance for the different approaches.
We conclude in Section \ref{sec:conclusions}.

%
%
\section{Methods} \label{sec:methods}
\subsection{Hydrodynamics and gravity}

We use {\tt FLASH}\footnote{\url{http://flash.uchicago.edu}} version 4  \cite{fryxell:2000, dubey:2009}
for all simulations carried out in this comparison. {\tt FLASH} is a publicly available, grid-based, parallel,
and multidimensional simulation framework to solve the compressible hydrodynamics equations.

The general setup is identical to what have been described in \cite{Couch:2013,
Couch:2014, oconnor:2018, Pan:2016, Pan:2018}.
Here we summarize the key features of our numerical approach.
We employ the directionally-unsplit hydrodynamics
(UHD) solver with third-order piecewise parabolic method (PPM) \cite{Colella:1984} spatial reconstruction,
a hybrid slope limiter, and a hybrid Riemann solver to calculate fluxes at zone interfaces.
The hybrid Riemann solver uses the HLLC Riemann solver in smooth regions
but reverts to using the more diffusive HLLE Riemann solver in zones
tagged as shocks to avoid odd-even oscillations \cite{quirk:1994}.

We use the multipole Poisson solver of Couch et al. \cite{Couch:2013}
with a maximum multipole value $l_{\rm max}=16$ for the calculation of self-gravity.
In order to approximate GR effects, We replace the monopole moment of the gravitational potential with an effective GR potential based on the case A implementation
described in \cite{marek:2006, oconnor:2018}.

We use 1D spherical and 2D cylindrical geometries in our simulations
with the PARAMESH (v.4-dev \cite{Macneice:2000}) library for adaptive-mesh-refinement (AMR).
We place the outer boundary of the domain at $10^4$~km and employ 9 levels of refinement.
We use 5 base AMR blocks in the radial dimension for both the 1D and
2D simulations, and 10 base AMR blocks along the cylindrical $z$-direction
in 2D.
Each AMR block contains 16 zones per dimension, giving a smallest zone width of $\sim$0.488\,km.
To save computation time, we reduce the maximum AMR level based on the distance from the center of the PNS,
enforcing an effective angular resolution of $< 0.52^\circ$.
We use a radial power law profile for density and velocity as outer boundary conditions
to approximate the stellar envelope rather than a pure outflow boundary condition which overestimates the
mass accretion flow and can affect the shock evolution at late times \cite{Couch:2013c}.

\subsection{Progenitor and nuclear equation of state}

We carry out our comparison using two progenitor models from
\cite{Woosley:2007} 
with zero-age main sequence masses of 15 and 20 $M_\odot$ (hereafter ``s15''
and ``s20'').
The structures
of these two progenitor models are quite different.  The s20 progenitor model
has a larger and denser silicon shell compared to the s15 model,
whose density declines much faster with radius in this region. This
leads initially to a larger mass accretion rate onto the PNS after
bounce in s20 when compared to  s15.
At the silicon-oxygen interface, the s20 model has a very strong
density gradient which gives a marked drop in the mass accretion
rate around $\sim$200\,ms after bounce.  Such a sharp
density drop is absent in s15, further distinguishing these
models. Finally, after the accretion of the silicon-oxygen interface the density structure in s20 is such that the mass accretion rate remains fairly constant,
but in s15 the mass accretion rate continues to slowly decrease.

We use the Steiner, Fischer, \& Hempel (SFHo) EoS,
which is tuned to fit, among other parameters, neutron star (NS) radius observations \cite{Steiner:2013}.
We consider two additional EoS: the Lattimer \& Swesty EoS
(with incompressibility $K= 220$~MeV, LS220) \cite{lattimer:1991},
and the Hempel \& Schaffner-Bielich (HS) EoS with the DD2 parameterization, HS(DD2) \cite{fischer:2014}.
Gauging the impact of different EoS gives us a point of comparison for interpreting the magnitude of the effect of different transport approximations.
Variants of the Lattimer \& Swesty EoS have been widely used in the CCSN simulation community since they first became available almost 30 years ago.
The LS220 EoS does not, however fulfill certain theoretical and experimental
nuclear physics constraints (see, e.g., \cite{hempel:2012,Kruger:2013}).
The HS(DD2) EoS, on the other hand, shows good agreement with nuclear experiment
about cluster formation properties \cite{fischer:2014}, but predicts larger NS radii than SFHo.
All three EoS have a maximum gravitational mass greater than 2\,$M_\odot$, as required by observations \cite{demorest:2010}.

%
%
\subsection{Neutrino transport} \label{sec:transport}

We compare three different neutrino transport implementations using the same hydrodynamics, gravity, and EoS, described above.
We use the two-moment explicit ``M1" closure method \cite{OConnor:2015},
the isotropic diffusion source approximation (IDSA) scheme \cite{Liebendorfer:2009},
and the advanced spectral leakage (ASL) method \cite{Perego:2014}.
In this section, we briefly described each method and highlight salient points of each relevant to our comparison effort.

%
%

\subsubsection{M1}

Our M1 transport scheme is a multidimensional, three-species,
energy-dependent, approximation to Boltzmann neutrino transport.
Instead of evolving the entire angle-dependent distribution function,
we only evolve the first two angular moments. The zeroth angular
moment represents the energy density of neutrinos within an energy
bin, while the first moment represents the momentum density of
neutrinos within an energy bin. We follow the formulation of
\cite{Shibata:2011a, OConnor:2015, oconnor:2018}. We simulate 12
energy groups, logarithmically spaced between 1\,MeV and 275\,MeV.  We
use opacities from {\tt NuLib} \cite{OConnor:2015}.  Briefly, these
include elastic scattering on nuclei and nucleons; charged current
absorption of electron type neutrinos and anti-neutrinos on nucleons
and electron type neutrinos on heavy nuclei; and thermal emission of
heavy-lepton neutrinos and anti-neutrinos from electron-positron annihilation and
nucleon-nucleon Bremsstrahlung.  These opacities and attendant corrections, including ion-ion correlations and the heavy nucleus form factor, are based on
\cite{Bruenn:1985, Burrows:2006a}. We neglect weak
magnetism corrections in order to more closely match the opacity sets used in IDSA and ASL.

The neutrino moment equations are solved using standard techniques
borrowed from hydrodynamics. In regions of low optical depth, the
evolution equations are hyperbolic and the spatial flux between grid
zones is determined using a Riemann solver.  In the high optical depth
limit, where the optical depth of a grid zone is greater than 1, we
change the flux determination from the Riemann solution to the
asymptotic diffusion limit fluxes.  To close the system of equations
we assume the M1 closure for the second moment \cite{OConnor:2015}. We
calculate the energy space fluxes (due to gravitational red shift and
velocity gradients) explicitly.  The neutrino-matter interaction source terms
are treated implicitly.

%
%
\subsubsection{IDSA}

A detailed description of IDSA is provided in \cite{Liebendorfer:2009} and \cite{Pan:2016}
Here, we briefly review the approach reiterating the
equations relevant for the present comparison.
In IDSA, the distribution function $f$ of
transported neutrinos is decomposed into a free-streaming component $f^s$
and a trapped component $f^t$. These two components
are evolved separately and linked by a diffusion source term $\Sigma$.
The diffusion source term is expressed as
\begin{equation} \label{eq_source}
\Sigma = \min \left\{ \max \left[\alpha + (j+\chi) \frac{1}{2} \int f^s d\mu, 0 \right] ,j \right\},
\end{equation}
where
\begin{equation} \label{eq_alpha}
\alpha = \nabla \cdot \left( \frac{-1}{3(j+\chi+\phi)} \nabla f^t \right),
\end{equation}
is a non-local diffusion scalar,
$j$ the emissivity, $\chi$ the absorptivity, $\phi$ the scattering opacity,
and $\mu$ the cosine of the angle between the neutrino propagation and the radial direction.
The trapped neutrino distribution $f^t$ is evaluated using Equation~\ref{eq_source},~\ref{eq_alpha}
and the transport equation
\begin{equation}
\frac{\partial f^t}{c~\partial t} = j - (j+\chi)f^t - \Sigma,
\end{equation}
assuming the spectral shape of the trapped component to be described by a Fermi distribution function.
The diffusion scalar $\alpha$ is solved by an explicit diffusion solver.
Once $f^t$ is determined, the net interaction rates $\mathcal{S}$ can be evaluated by
\begin{equation}
\mathcal{S} = \frac{\partial f^t}{c\partial t} + \Sigma - (j +\chi) \frac{1}{2} \int f^s d \mu,
\end{equation}
and hydrodynamics quantities are updated by:
\begin{equation}
\frac{\partial Y_e}{c \partial t} = - \frac{m_b}{\rho} \frac{4\pi c}{(hc)^3} \int \left( \mathcal{S}_{\nu_e} - \mathcal{S}_{\bar{\nu}_e} \right) E^2 dE,
\end{equation}
\begin{equation}
\frac{\partial e}{c \partial t} = - \frac{m_b}{\rho} \frac{4\pi c}{(hc)^3} \int \left( \mathcal{S}_{\nu_e} + \mathcal{S}_{\bar{\nu}_e} \right) E^3 dE -Q_x,
\end{equation}
where $\rho$ is the matter density, $m_b$ the baryon mass and $h$ the Planck constant.
$Q_x$ is the cooling provided by
$\mu$ and $\tau$ neutrinos which is modeled by a grey leakage scheme \cite{Hannestad:1998, rosswog:2003}.
In our current IDSA solver, we only consider the transport of electron flavor neutrinos and anti-neutrinos.
The streaming neutrino distribution $f^s$ and streaming neutrino flux for the next step can be calculated
from the neutrino net interaction rates and the streaming transport equation.
To couple the trapped neutrino component with matter, we introduce the trapped neutrino fraction,
\begin{equation} \label{eq_ynu}
Y^t_\nu = \frac{4 \pi m_b}{\rho (hc)^3} \int f^t E^2 dE d\mu,
\end{equation}
and energy,
\begin{equation} \label{eq_znu}
Z^t_\nu = \frac{4 \pi m_b}{\rho (hc)^3} \int f^t E^3 dE d\mu \, .
\end{equation}
The quantities $\rho Y^t_\nu$ and $(\rho Z_\nu^t)^{3/4}$ are advected with the fluid. 
The scaling used for the trapped neutrino energy fraction  
ensures the inclusion of compressional heating from trapped neutrinos via neutrino pressure.
These contributions have been shown to be crucial in CCSN simulations by \cite{Mezzacappa:1993}.
Note that the current IDSA solver does not include any GR corrections to the transport equations.

The IDSA solver was first implemented in 1D
coupled to the {\tt AGILE} hydrodynamics code \cite{Liebendorfer:2009}
and compared with the {\tt AGILE-BOLTZTRAN} code \cite{Liebendorfer:2004} in the Newtonian limit.
The latter solves the full Boltzmann transport equation.
Good agreement of neutrino fluxes and spectra between IDSA and full Boltzmann transport was found in \cite{Liebendorfer:2009},
but IDSA leads to a slightly larger maximum shock radius ($\sim$10-20\%) and a faster shock contraction.

To extend the IDSA solver to multiple dimensions, one could either
solve for the diffusion scalar $\alpha$ in multiple dimensions,
but keep the streaming component isotropic \cite{Pan:2016}, or
implement the IDSA with a `ray-by-ray plus' approach
\cite{Suwa:2013, Takiwaki:2014, Kotake:2018}.
In the latter case, the domain is decomposed in several radial directions, along which
the transport problem is solved separately as in spherical symmetry, but neutrino
quantities can be still advected in multiple dimensions.
In this paper, we implement the IDSA solver in {\tt FLASH} with the former approach,
keeping the diffusion scalar in multi-D but solving
the streaming component isotropically.
12 energy bins that are logarithmically spaced from 3 to 200~MeV are used in the IDSA solver.
We use neutrino rates for the emission, absorption, and scattering of neutrinos off neutrons,
protons and nuclei from \cite{Bruenn:1985} and nucleon-nucleon Bremsstrahlung from \cite{Hannestad:1998}.
We note that our IDSA implementation is different to the IDSA implementations of
\cite{Suwa:2013, Takiwaki:2014, Kotake:2018}. For instance, 
\cite{Takiwaki:2014, Kotake:2018} include heavy lepton neutrinos in the IDSA solver and also 
account for inelastic neutrino-electron scattering.

%
%
\subsubsection{ASL}
The Advanced Spectral Leakage (ASL) method \cite{Perego:2016, Perego:2014} is a
three-species approximate neutrino treatment designed to model neutrinos in the
context of core-collapse supernovae and compact binary mergers.
It is based on previous gray leakage schemes \cite{Ruffert:1998, rosswog:2003, OConnor:2010}, but in addition it carries spectral
information on discretized neutrino energies and models trapped neutrino  components.
The spectral particle and energy emission rates are computed as a smooth interpolation between local production and
diffusion rates. The former are the relevant rates in optically thin conditions, while the latter
are computed based on timescale arguments and become relevant in optically thick regions.
The modeling of the neutrino trapping at high densities is achieved by
the solution of advection equations for $Y_{\nu}$ and $Z_{\nu}$ (Equations (\ref{eq_ynu}) and (\ref{eq_znu}))
in analogy with the IDSA. The reconstruction of the neutrino distribution
functions from $Y_{\nu}$
is based on weak and thermal equilibrium arguments. Neutrino absorption rates in optically
thin conditions are also included, based on the calculation of the neutrino densities of the free streaming
neutrinos. ASL has been implemented in several hydrodynamics codes, both Eulerian and Lagrangian, including
{\tt AGILE} \cite{Liebendorfer:2002} in spherically symmetry, FISH \cite{Kappeli:2011} and SPHYNX
\cite{Cabezon:2017} in 3D.

At the moment, ASL includes neutrino emission and absorption on free nucleons and
nuclei \cite{Bruenn:1985}, neutrino scattering off nucleons and nuclei in the elastic approximation \cite{Bruenn:1985},
electron-positron annihilation \cite{MezzacappaMesser:1999}, and
nucleon-nucleon Bremsstrahlung \cite{Hannestad:1998}.

In our current FLASH implementation, we evaluate the neutrino emissivities and opacities
at 12 energy bins logarithmically spaced between 3 and 300~MeV. The ASL
free-streaming component follows the ray-by-ray implementation of the gray
leakage scheme available in FLASH \cite{Couch:2014}, where we map the
hydrodynamic grid quantities onto a set of radial rays and split the
multi-D problem into several 1D calculations. In 2D, the ray-grid consists of
$1000$ radial zones, linearly spaced up to $150$ km and logarithmically spaced
up to $3000$ km, and 37 uniform angular zones providing a
resolution of $4.9^{\circ}$.\\
On each ray, we compute the energy-dependent optical depths and the spectral neutrino
density.
Locally, we map these quantities back to the hydrodynamics grid and evaluate the
local source terms. The fluid energy and electron fraction, as well as the
neutrino particle and energy densities, are then updated in an explicit way.
The implementation presented in \cite{Perego:2016} is purely
Newtonian. Due to the inclusion of an  effective GR gravitational potential,
we have developed a relativistic extension that takes into account the most relevant
relativistic effects (e.g., relativistic Doppler effect and gravitational redshift) in the
propagation of the free streaming neutrinos in optically thin conditions,
see \ref{Appendix:ASL_relativistic_extension}.
Neutrino stress in the momentum equation and from free streaming particles is not taken 
into account in the current implementation. 

ASL contains three free parameters that require calibration:
$\alpha_{\mathrm{blk}}$, $\tau_{\mathrm{cut}}$, and $\alpha_{\mathrm{diff}}$.
As outlined in \cite{Perego:2016}, $\alpha_{\mathrm{blk}}$ affects the total luminosity and the heat deposition, $\tau_{\mathrm{cut}}$
the neutrino energy, and $\alpha_{\mathrm{diff}}$ the PNS cooling rate. In \cite{Perego:2016} these parameters
are calibrated in 1D against detailed Boltzmann transport, using the \texttt{AGILE-BOLTZTRAN} code \cite{Liebendorfer:2002}, a 15 $M_{\odot}$ 
zero-age main sequence mass progenitor
\cite{Woosley:2002}, and the LS220 EoS. The
calibrated (standard) values were $\alpha_{\mathrm{blk}} = 0.55$,
$\tau_{\mathrm{cut}} = 20$, and $\alpha_{\mathrm{diff}}=3+2X_h$,
where $X_h$ is the mass fraction of heavy nuclei. Since the implementation of
ASL used in this comparison also contains GR corrections in contrast to the
original implementation, we have repeated the calibration using the
\texttt{FLASH-M1} code in 1D, the s20 progenitor and the SFHo EoS as reference
case. We have obtained $\alpha_{\mathrm{blk}} = 0.5$, $\tau_{\mathrm{cut}} = 15$,
and $\alpha_{\mathrm{diff}}=3+2X_h$, comparable with the standard
parameter set presented in \cite{Perego:2016}.
We have tested that differences between models employing the standard and the
recalibrated parameter sets do not qualitatively change the simulation
outcome for the calibration setup, but the original parameters lead to
undesirable quantitative discrepancies for a detailed comparison.
For instance, the maximum shock radius is reached about 10~ms later.

%
%
\subsection{Initial condition}

We evolve the s15 and s20 progenitors from core collapse to $\sim$15\,ms post-bounce
using {\tt GR1D} \cite{OConnor:2010, OConnor:2015} and then remap the
simulations to {\tt FLASH}.
{\tt GR1D} employs the same M1 scheme as in {\tt FLASH-M1}
but additionally includes inelastic process that are not included in {\tt FLASH-M1},
{\tt FLASH-IDSA}, and {\tt FLASH-ASL} (hereafter, M1, IDSA, and ASL).
This post-bounce remapping approach is similar to that employed in \cite{oconnor:2018}.

Restarting a {\tt GR1D} simulation with M1 is straightforward,
since M1 and {\tt GR1D} share identical variables and inelastic processes
are subdominant during the accretion phase after core bounce.
However, in IDSA and ASL only trapped neutrinos are
advected with the fluid.
We, therefore, have to decouple trapped neutrinos from the total neutrinos in
the initial conditions obtained from {\tt GR1D}.
We assume that the neutrino flux in M1 is purely from the free-streaming neutrinos
and use the flux factor suggested in \cite{Liebendorfer:2004},
where it was assumed that all neutrinos with
a given energy are isotropically emitted at their last scattering neutrino sphere.
Therefore,
\begin{equation}
\frac{1}{2} \int f^s(E) d\mu = \frac{2 \mathcal{F}(E)}{1 + \sqrt{1-\left( \frac{R_\nu(E)}{\max \left(r,R_\nu(E)\right)}\right)^2}},
\end{equation}
where $\mathcal{F}(E)$ is the neutrino flux at energy $E$, and $R_\nu(E)$ is
the corresponding neutrino sphere. $R_\nu(E)$  is determined from
energy-dependent opacities and is defined as the 
radius where the energy-dependent optical depth becomes $2/3$.
Once we know the distribution function of the free-streaming neutrinos,
the distribution function of the trapped neutrinos is simply
$f^t(E) = \max(f(E) - f^s(E),0)$, and the neutrino fraction and energy can be
calculated by using Equations~(\ref{eq_ynu}) and (\ref{eq_znu}).

The comparison in the following sections are done with 12 energy bins in all three transport schemes,
but it should be noted that the maximum energy bin in IDSA (200~MeV) is lower than 
the maximum energy bin in M1 (275~MeV) and ASL (300~MeV).
We have tested each transport scheme with a varying number of energy
bins (from 12 up to 20 bins) in both 1D and 2D.
No significant differences (other that stochastic variations for the
2D simulations) were found. 
Furthermore, we note that the opacity contributions from various neutrino interactions are nearly identical
in the IDSA and ASL schemes and can be found in the public version of the Agile-IDSA \cite{Liebendorfer:2009}. 
The M1 scheme includes the same set of neutrino interactions but packed in the open-source {\tt NuLib} library.
We do not expect significant differences in our comparison study due to these opacity sets,
since they implement similarly neutrino interactions. However, 
different opacity sets invoking different neutrino interactions can make appreciable differences as discussed in \cite{Lentz:2012, Kotake:2018}.

%
%

\section{Results} \label{sec:results}

%
%

\begin{figure}
\includegraphics[width=1.0\textwidth]{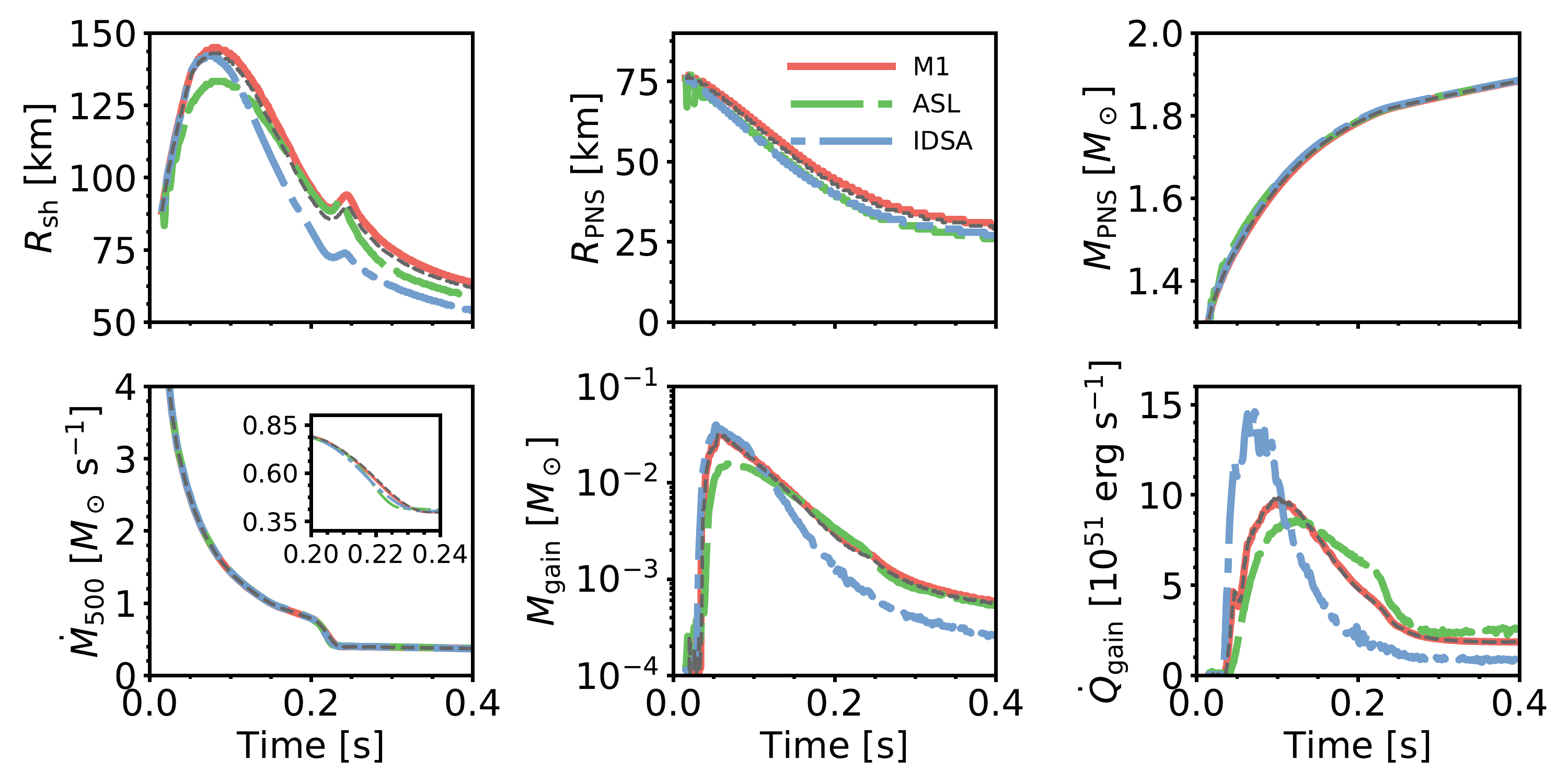}
\caption{Time evolution of several shock radius, PNS radius, PNS mass, mass accretion rate,
mass in the gain region, and heating in the gain region for the simulations of s20 progenitor
with SFHo EoS. Different color represents simulation with different neutrino transport scheme.
The gray thin line shows the same M1 run but with weak magnetism corrections, 
which is directly comparable to \cite{OConnor2018b}.
\label{fig:1d}
}
\end{figure}

\subsection{Transport comparison in spherical symmetry}
\label{s.1d}

We perform a series of 1D spherically symmetric simulations with M1, IDSA, and ASL.
We begin with a comparison of s20 with the SFHo EoS.
Figure~\ref{fig:1d} shows the time evolution of shock radius, PNS
radius and mass, mass accretion rate, and the mass and net neutrino
heating rate in the gain region obtained with the different transport
schemes.
The mass accretion rate (measured at 500\,km radius) and PNS mass are nearly identical
in all three transport schemes, verifying that the initial conditions, and the hydrodynamic and gravity solvers
are consistent in all three schemes.
M1 restarts smoothly from the GR1D and we do not observe any noticeable effects due to relaxing the model on the FLASH grid.
IDSA takes about 5~ms to relax the GR1D quantities and shows some small oscillations on the shock radius and neutrino quantities.
ASL takes longer ($\sim 30$~ms) to relax the GR1D model and a more
pronounced oscillation of the PNS radius (likely due to the lack of
neutrino pressure) is observed 
(see Figure~\ref{fig:1d}). Since the transition from GR1D to FLASH-M1 is smooth, 
we attribute these early oscillations seen in IDSA and ASL to
differences in the treatment of neutrino transport.

\begin{figure}
\centering
\includegraphics[width=1.0\textwidth]{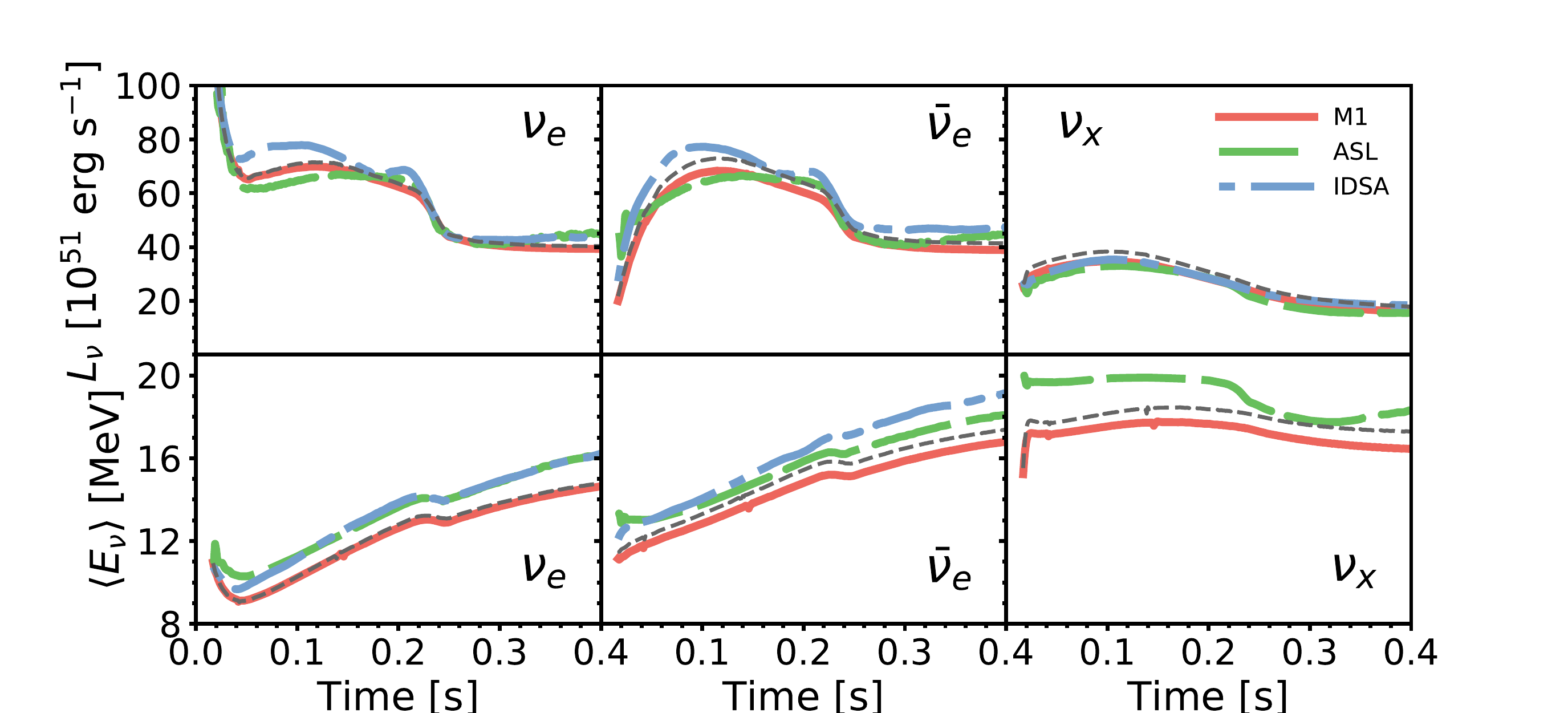}
\caption{Time evolution of of neutrino luminosities (top row) and mean energies (down row)
for the simulations of s20 progenitor with SFHo EoS.
Different color represents simulation with different neutrino transport scheme.
The gray thin line shows the same M1 run but with weak magnetism corrections, 
which is directly comparable to \cite{OConnor2018b}.
\label{fig:1d_neutrino} }
\end{figure}

\begin{figure}
\centering
\includegraphics[width=1.0\textwidth]{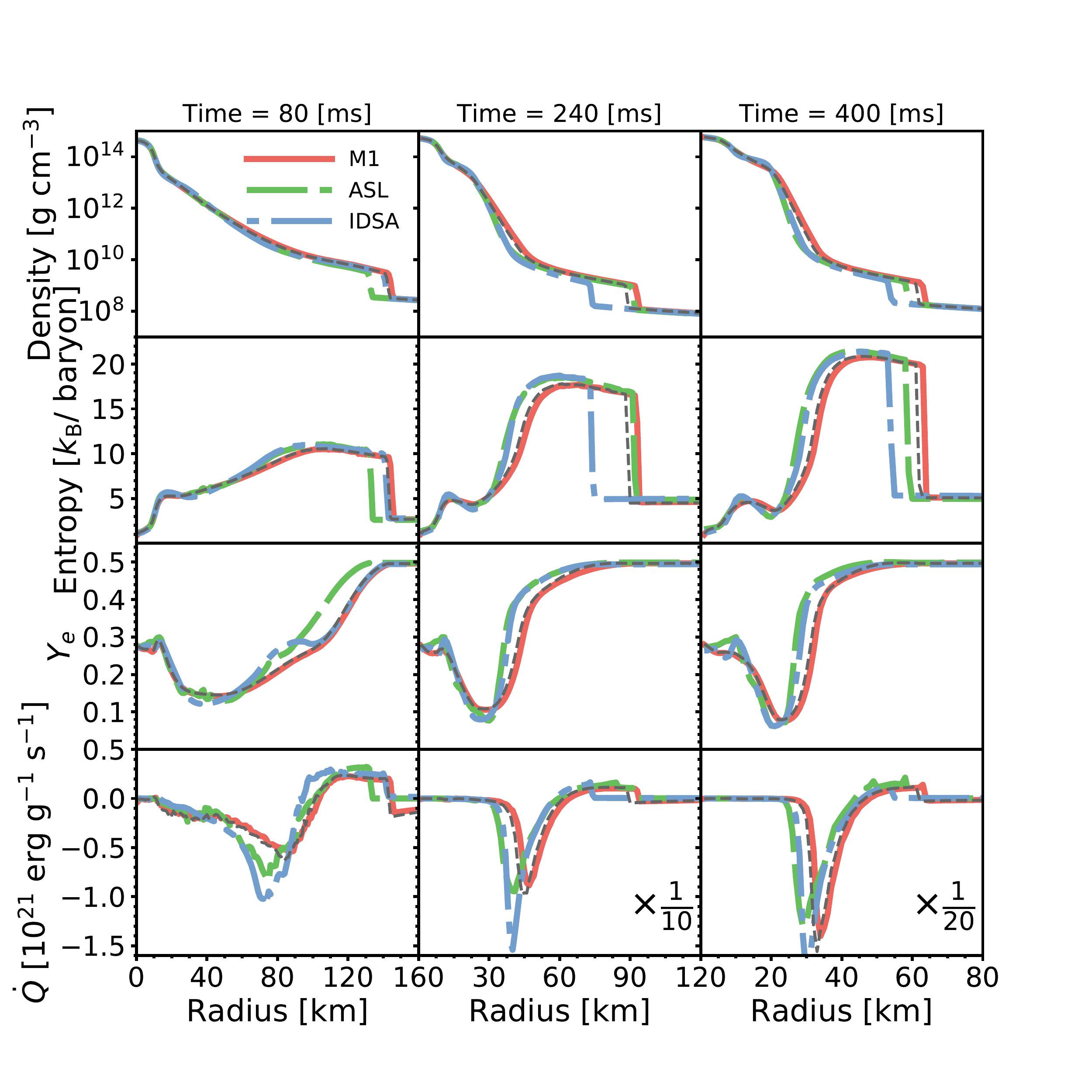}
\caption{1D radial profiles at different post bounce times for the simulations of s20 progenitor with SFHo EoS.
Different color represents simulation with different neutrino
transport scheme.
The gray thin line shows the same M1 run but with weak magnetism corrections, 
which is directly comparable to \cite{OConnor2018b}.
\label{fig:1d_radial}}
\end{figure}

During the first $\sim$100\,ms post-bounce, M1 and IDSA show a very similar shock radius evolution
that peaks at $\sim$145\,km, while the ASL run has a relatively smaller shock radius
evolution and peaks at $\sim$135\,km. At $\sim$15\,ms, the shock radius in the ASL simulation becomes
comparable to the M1 simulation, and after this time the evolution is
similar.
On the other hand,
the IDSA run gives a $\sim$15\% smaller shock radius at $\sim$200\,ms post-bounce,
but this difference becomes smaller at late times.
All three schemes give a similar PNS radius and mass,
but the M1 scheme has a slightly larger PNS radius,
which might be due to the different treatment of the heavy-lepton neutrinos
Figure~\ref{fig:1d_neutrino} shows the time evolution of neutrino luminosity and mean energy
obtained with the three transport schemes.
When the shock has stalled at $\sim$80\,ms,
the IDSA (ASL) run has the highest (lowest) electron neutrino luminosity
$\sim$80$\times 10^{51}$\,erg\,s$^{-1}$ ($\sim$60$\times 10^{51}$\,erg\,s$^{-1}$) among the three schemes,
and the M1 run lies between the IDSA and ASL.
The same trend can be seen in the electron anti-neutrino luminosity.
The higher luminosity in the IDSA  simulation is due to a larger contribution to the steaming neutrinos at radii outside the neutrinosphere.
The IDSA and M1 show similar shock radius and PNS radius (which is approximately the radius of the neutrinosphere),
resulting in a similar gain radius, but IDSA has a higher heating rates at early time and a lower heating rates at late time.
We note that IDSA has a second bump
on its electron neutrino luminosity at $\sim$200\,ms.
This feature does not exist in either the M1 or ASL runs,
but the electron neutrino luminosity after the second bump in the IDSA matches
with the M1 and ASL runs at $t > 200$~ms.
A transition between two limiting cases in the IDSA diffusion solver
in Equation~\ref{eq_source} is a possible origin of this feature.

Both IDSA and ASL show slightly higher electron neutrino and electron anti-neutrino mean energy than M1
in the first 200~ms. The difference grows to $\sim$15\% after $\sim$200\,ms post-bounce
(see Figure~\ref{fig:1d_neutrino}). The $\mu/\tau$ neutrino mean
energy in ASL is usually $\sim 15\%$ higher than in M1. This excess reduces to
$\sim 10\%$ only after the drop in the accretion rate due the progenitor shell
interface. This larger excess reveals the challenge for leakage schemes to model
extended scattering atmospheres. We note that the leakage solver
for $\mu/\tau$ neutrinos in IDSA does not track the mean
energies and therefore these are not plotted in Figure~\ref{fig:1d_neutrino}.

Figure~\ref{fig:1d_radial} shows the radial profiles of density, entropy, electron fraction,
and heating rates of the three transport schemes at different post-bounce times.
By $\sim$240\,ms, all schemes have developed 
a negative entropy gradient just below the PNS radius. The simplified treatment of the diffusive 
regime in leakage schemes prevents an effective transport and redistribution of the heat inside the 
optically thick PNS. As a consequence, the dominant heavy-flavor cooling at the PNS surface produces a
more pronunced entropy gradient in ASL and IDSA, compared to M1.
The negative heating rates of M1 outside of the shock front at 80~ms post-bounce
are due to the exchange of momentum from streaming neutrinos.
It should be noticed that ASL and IDSA only include neutrino compressional heating
from trapped neutrinos, but no neutrino stress from free streaming neutrinos.
During the entire post bounce evolution, M1 has a larger radial extent
than IDSA and ASL, which was also noted in the PNS radii panel of Figure~\ref{fig:1d}.
Apart from this, at 240 and 400~ms, all three schemes give very consistent radial profiles,
except the IDSA run has a smaller shock radius at 240~ms. 

As a final note, we compare and contrast our results to the recent
work of \cite{OConnor2018b}.  In the referenced comparison work,
several neutrino transport codes are compared in 1D using very similar
conditions to those used here, the only major difference is that in
this work here we neglect weak magnetism and recoil corrections. In
that work, differences between various quantities among all the codes
ranged from 5\% to 15\%, with some excursions upwards to 50\% for some
select quantities at late times (e.g. neutrino energies and net heating in the gain region).
To link our results to \cite{OConnor2018b}, we include the FLASH-M1 run 
with weak magnetism and recoil corrections as the thin gray lines in Figure~\ref{fig:1d} 
and Figure~\ref{fig:1d_neutrino}. A higher electron type neutrino and antineutrino luminosity in the IDSA can be seen 
in the 3DnSNe-IDSA code \cite{Kotake:2018} as well.

%
%

\subsection{A comparison with different EoS}

\begin{figure}
\centering
\includegraphics[width=1.0\textwidth]{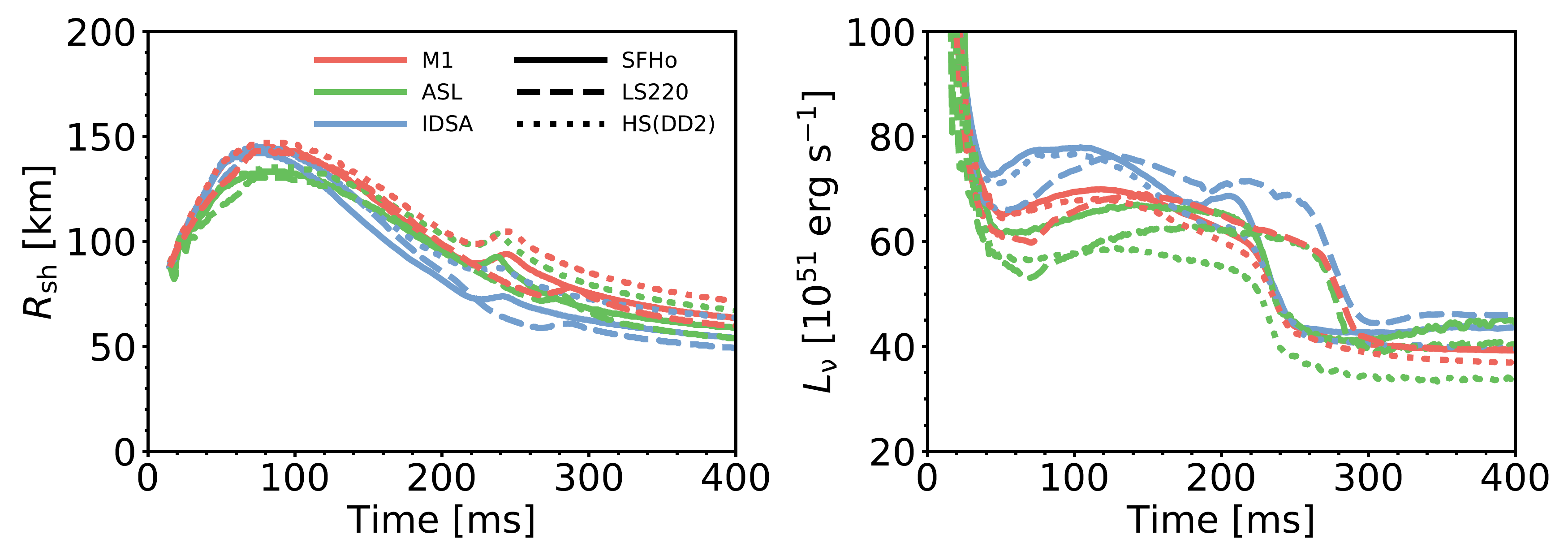}
\caption{Time evolution of several shock radius and electron neutrino luminosity for the simulation of s20 progenitor.
Different line style represents different EoS and different color
shows simulations with different neutrino transport schemes. 
\label{fig:1Deos}}
\end{figure}

Recent CCSN simulations suggest that the EoS could have impact
on the explodeability \cite{Couch:2013c, Suwa:2013},
on SASI activity \cite{Kuroda:2016},
on the dynamics of stellar-mass black hole formation \cite{Pan:2018}, 
and on gravitational wave and neutrino signals \cite{Kuroda:2016, Richers:2017, Pan:2018, Morozova:2018}.
In order to disentangle the effects of the EoS from the neutrino transport methods,
we perform 1D simulations of the
s20 progenitor for all the three transport methods with two additional EoSs: LS220 and HS(DD2).
The time evolution of shock radius and electron neutrino luminosity can be seen
in Figure~\ref{fig:1Deos} for the three transport schemes we consider.
The LS220 runs have a later drop in the neutrino luminosity from about $\sim$250\,ms to $\sim$300\,ms
due to a different treatment of the low density EoS that causes a different mass accretion evolution.
The simulations using the HS(DD2) EoS give the largest shock radius, followed by simulations using SFHo and LS220 EoS.
The runs with SFHo (LS220) EoS have the highest (lowest) electron neutrino luminosity, respectively.
The neutrino luminosity with the HS(DD2) EoS is slightly lower than
that with SFHo. 

All three transport schemes show the same trends while varying the EoS,
suggesting that the usage of different EoS has a lower impact than the usage of different transport schemes.
Therefore, the differences we discussed in the previous section do not depend on the specific choice of the EoS.

%
%

\subsection{Transport comparison in cylindrical symmetry}
\label{sec:2Dcomparison}

\begin{figure}
\centering
\includegraphics[width=\textwidth]{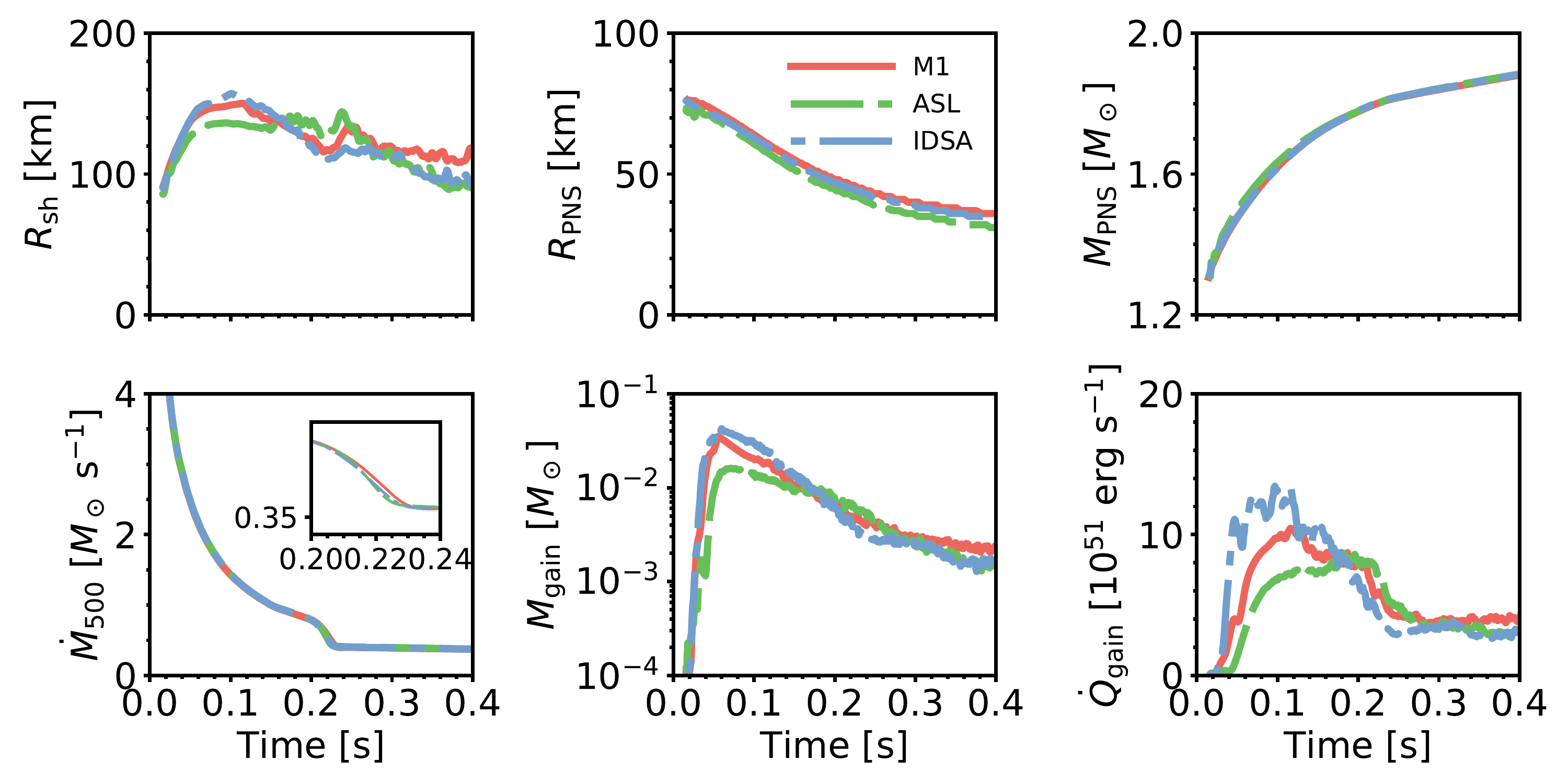}
\caption{Time evolution of several shock radius, PNS radius, PNS mass, mass
accretion rate, mass in the gain region, and heating in the gain region for the
simulations of s20 progenitor with SFHo EoS in 2D. Different color represents
simulation with different neutrino transport scheme.
\label{fig:2D}}
\end{figure}

In this section, we extend our comparison to multiple dimensions by
comparing the three transport schemes via 2D cylindrically-symmetric
simulations. In Figure~\ref{fig:2D}, we show the same quantities for
our 2D simulations as we have shown for the 1D simulations in
Figure~\ref{fig:1d}.  The overall behavior is very similar to 1D
until about $\sim$100\,ms when convection begins to take hold in the
gain region, breaking the spherical symmetry as visible by stronger
shock oscillations, and non-zero anisotropic velocities. Up to
$\sim$400\,ms post bounce, none of the three models shows signs of
incipient explosions.

Figure~\ref{fig:2D} reveals that both the PNS mass and accretion
rate evolve similarly for all treatments since they are essentially
determined by the underlying progenitor structure and gravity,
neither of which is strongly impacted by the neutrino transport
scheme or dimensionality. This is consistent with our results in
1D (Section \ref{s.1d}).  Once spherical symmetry is broken and convection
becomes non-linear (after $\sim$100\,ms) several of these displayed
quantities begin to deviate from the 1D results. The first
noticeable deviation is in the shock radius (top left panel of
Figure~\ref{fig:2D}), which reaches roughly the same maximum radius
($\sim$135\,km for ASL, $\sim$150\,km for M1 and $\sim$155\,km for
IDSA), but then has a much slower decline. When the silicon/oxygen
interface accretes through the shock at $\sim$220\,ms, the shock
radii are between $\sim$110-130\,km, which is $\sim$30-40\,km more
than the value at the corresponding time in 1D. This is due to the
additional dynamical pressure support and dissipation from the turbulent motions behind the
shock \cite{murphy:2013, couch:2015a, mabanta:2018}. In the bottom-center and bottom-right panels of Figure
\ref{fig:2D}, we show the mass and the neutrino heat deposition rate
in the gain region, respectively.  These quantities further show the
qualitative effect of multidimensional dynamics on the CCSN central
engine.  Compared to the analogue quantities for the 1D cases in
Figure \ref{fig:1d}, we notice a slower decrease of the mass in gain
region and an increased heat deposition at later times. Both of
these are a result of, and also contribute to, the presence of
aspherical flows in the gain region and the increased shock radius. Lastly, we
note that as seen in Figure \ref{fig:2D}, the PNS radius is decreasing at a
slower rate in the 2D simulations compared to the equivalent 1D simulations. 
This is due to the presence of PNS convection.
  
\begin{figure}
\centering
\includegraphics[width=1.0\textwidth]{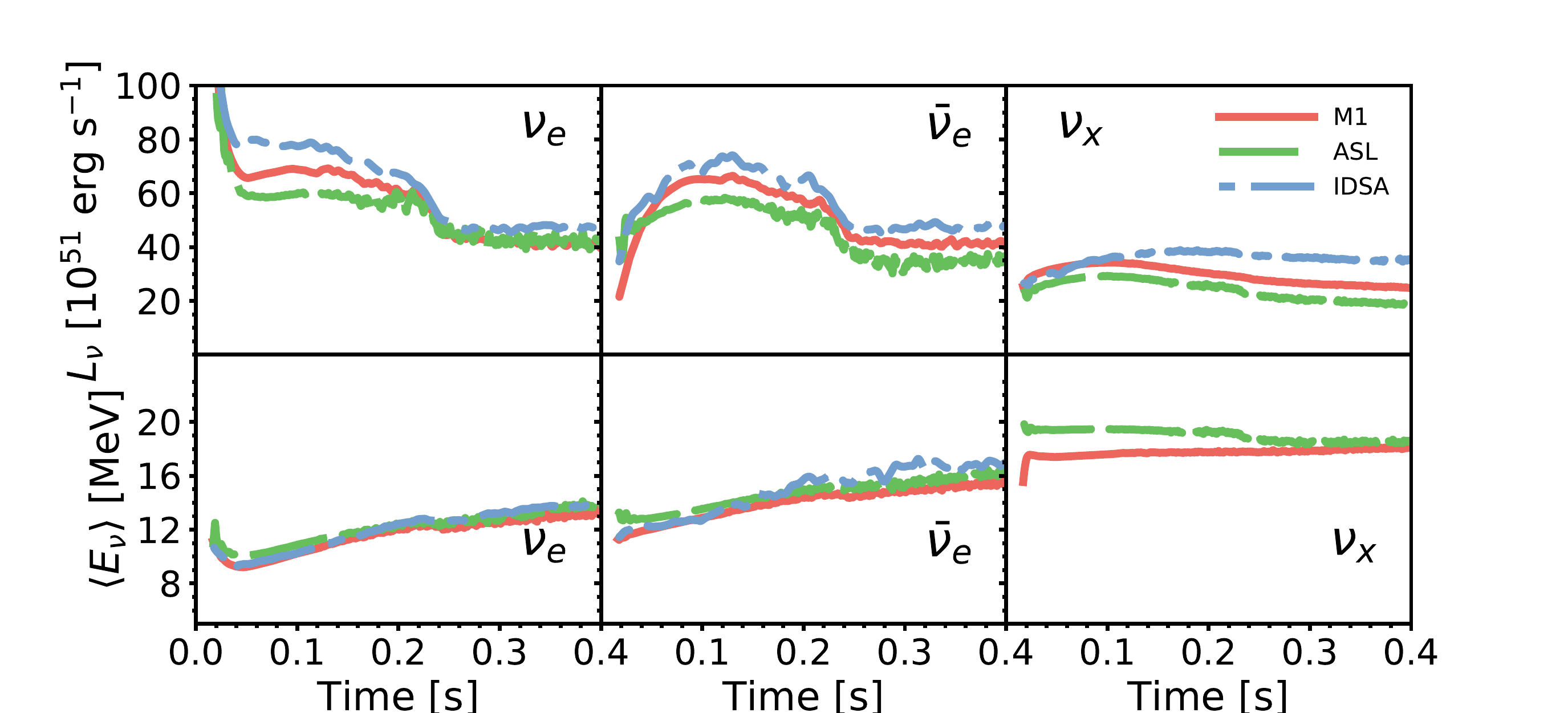}
\caption{Time evolution of of neutrino luminosities (top row) and mean energies
(down row) for the simulations of s20 progenitor with SFHo EoS in 2D. Different
color represents simulation with different neutrino transport
scheme.
\label{fig:2D_neutrino}}
\end{figure}

Figure~\ref{fig:2D_neutrino} shows the evolution of neutrino luminosities and
mean energies for the 2D simulations. Compared to the 1D simulations, the
non-spherical accretion of turbulent material onto the PNS leads to variable
signals on small timescales.  This is most evident at later times, after
convection has fully developed, and in the electron neutrino and anti-neutrino
signals, which are emitted closer to the material in the convection zone. The
heavy-lepton neutrinos originate from deeper inside in the gravitational well
where the fluid motions are calmer. In 2D, convection inside the PNS increases
the heat transfer from the opaque center to the surface where neutrino cooling
is more efficient. This results in higher luminosities for the representative
$\nu_x$ species compared to 1D models (see the top-right panel in
Figure~\ref{fig:1d_neutrino} and Figure~\ref{fig:2D_neutrino}). The M1
simulation has less of an enhancement, consistent with the milder PNS convection
discussed below.

\begin{figure}
\centering
\includegraphics[width=1.0\textwidth]{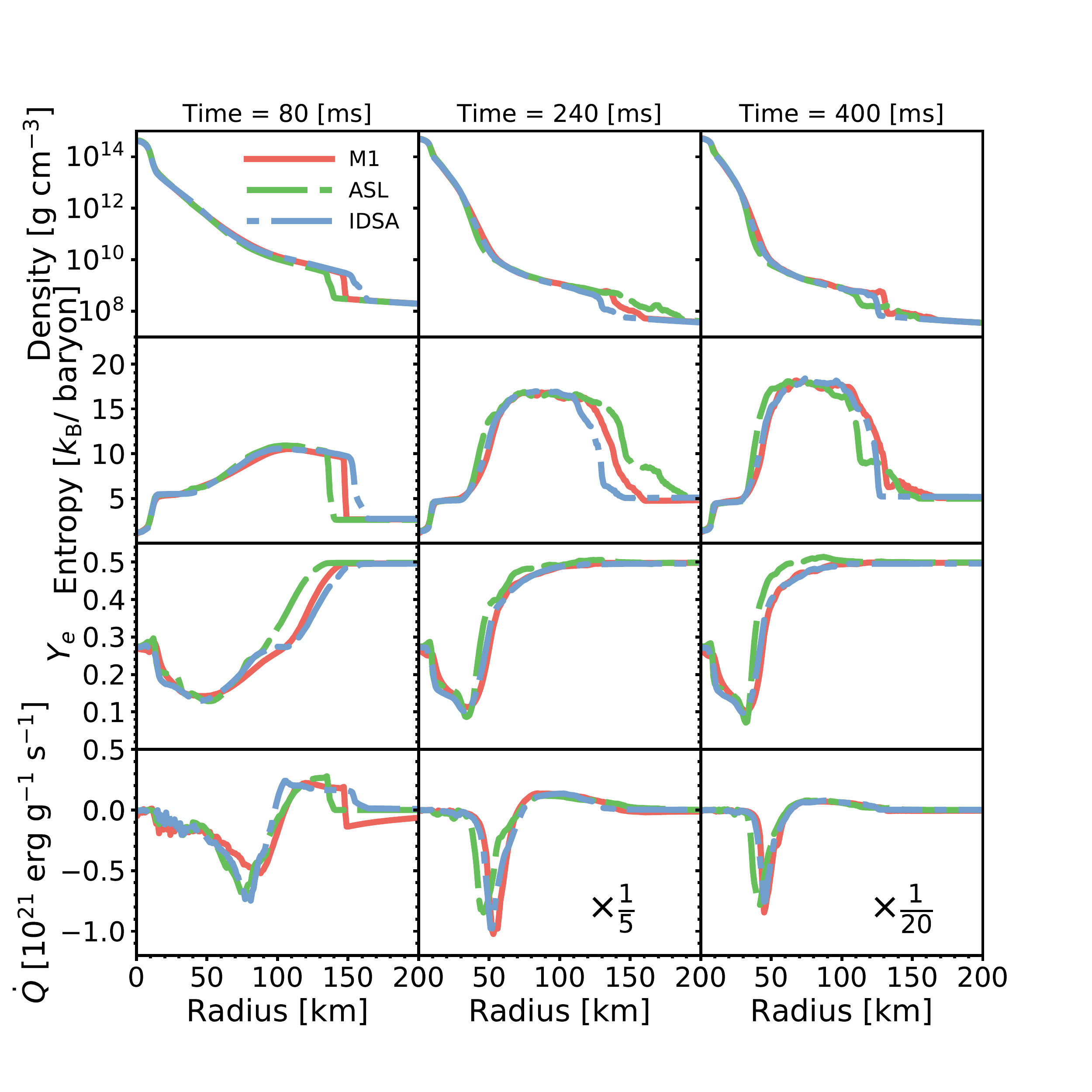}
\caption{Angular averaged radial profiles at different post bounce times for
the simulations of s20 progenitor with SFHo EoS in 2D. Different color
represents simulation with different neutrino transport scheme.
\label{fig:2D_radial}}
\end{figure}

In order to understand where the differences among the neutrino treatments
occur, we consider radial profiles of angular averages at different
simulation times (see Figure~\ref{fig:2D_radial}). These profiles confirm that
also in 2D the PNS is very similar for the different schemes. 
However, in contrast to the 1D case, the negative entropy and lepton gradients trigger
PNS convection, which leads to a flatter entropy profile below the PNS radius.
The density and entropy per nucleon (first two rows in 
Figure~\ref{fig:2D_radial}) compare well at small radii, where
all of the matter is shocked in each transport scheme. The differences that
do arise at small radii are consistent with variations of the PNS radius. The
ASL simulation shows the most compact PNS, while the PNS radius in M1 and IDSA are
slightly larger. The radial profiles of density and entropy can differ
substantially below the shock and inside the gain region where the angular
averages contain both shocked and unshocked matter at various percentages for
the different schemes. A direct comparison in this regime is less straightforward. 

The neutrino radiation is chiefly coupled to the matter via energy and
electron fraction source terms. In the last two rows of
Figure~\ref{fig:2D_radial} we show quantities related to these source terms, 
i.e. the electron fraction (second last row), and the rate of energy exchange
between the matter and the neutrinos (bottom row; negative values
means the matter is losing energy to neutrino interactions).  In
general, the matter begins to deleptonize after it accretes through
the shock and dissociates into free neutrons and protons. The lower
shock radius for ASL at 80\,ms accounts for the difference in $Y_e$
seen there.  At late times ASL tends to exhibit larger electron
fractions, even greater than 0.5, close to the shock position. The
$Y_e$ in this regime is set by the deleptonization rate, but also via
the neutrino heating. In the ASL simulation, the marginally but systematically 
larger electron neutrino luminosity enhances the rate of conversion of neutrons 
into protons inside the heating region. Moreover, the ray-by-ray scheme 
tends to enlarge the relative differences between the electron neutrino and 
electron anti-neutrino spectra as seen in the luminosities in 
Figure~\ref{fig:2D_neutrino}.

The neutrino energy source term reveals the location and 
strength of the neutrino interaction, see the last row in
Figure~\ref{fig:2D_radial}. Especially during the shock expansion
at 80~ms, IDSA and ASL show a very similar cooling signature below the
gain radius which is located at about 100 km. As in 1D, M1 cools less
inside this region. Above the gain radius up to the shock radius, M1
and ASL show a similar heating signature where IDSA deposits slightly
more heat. At radii above the shock ASL and IDSA have a vanishing
neutrino energy source term, but M1 also takes the neutrino pressure
work on the in-falling matter into account.
Furthermore, we note that comparing to 1D (Fig.~\ref{fig:1d_radial}) 
where M1 shows a larger radial extent in all given profiles (e.g.~the rising 
entropy between 30--60~km at 240~ms), all schemes show a closer agreement in 
the radial extent of their equivalent 2D profiles (Fig.~\ref{fig:2D_radial}).
Here, the strong PNS convection in ASL and IDSA lessens the difference
and leads to an apparent equalization of the PNS radii among the schemes.

\begin{figure}
\centering
\includegraphics[trim={0 0 175 0},clip,width=\textwidth]{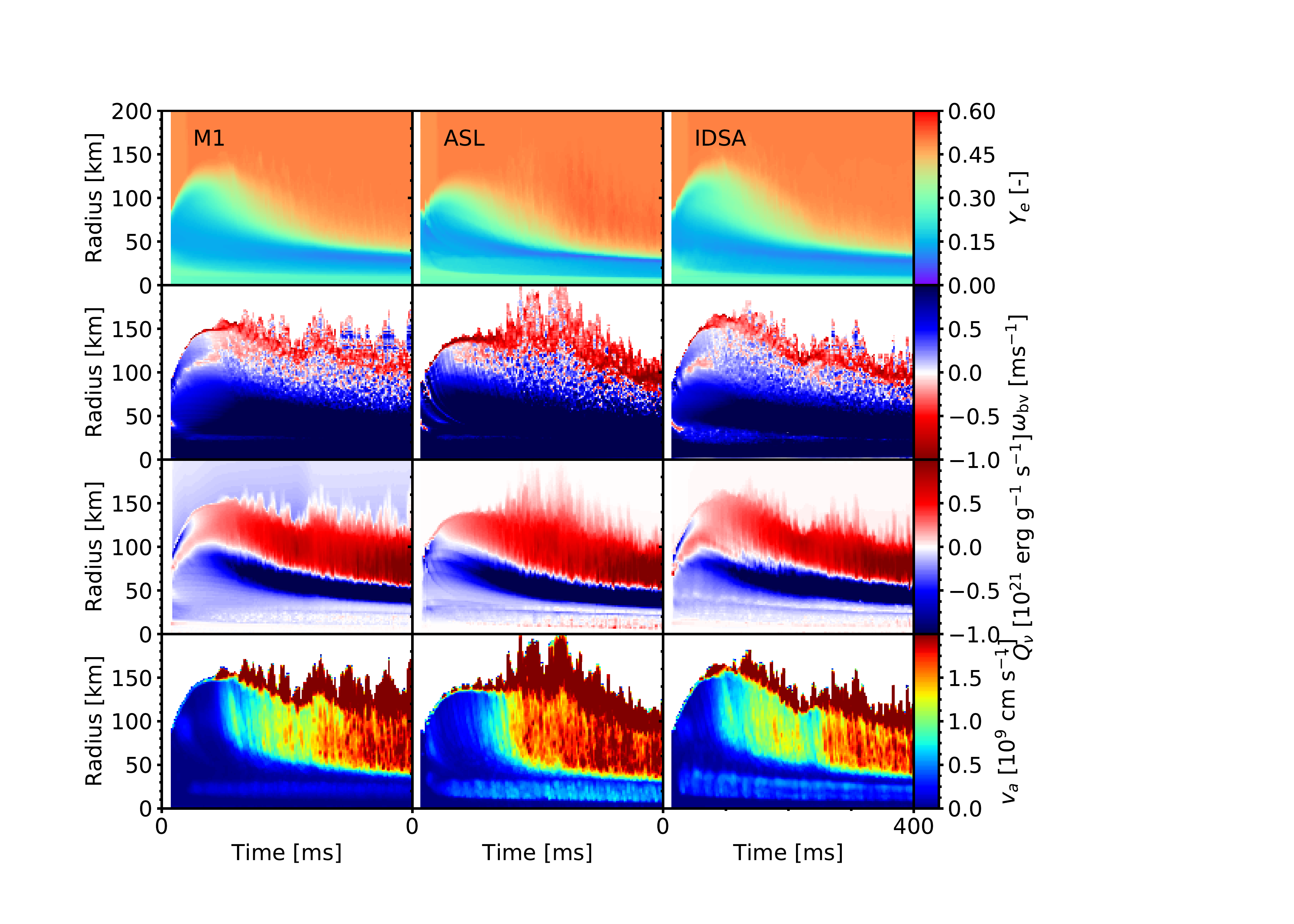}
\caption{Time evolution of radial averaged profiles. Each column represents
the data of one of the schemes. The first row shows the electron fraction
$Y_e$, the second row are the profiles of the Brunt-V\"ais\"al\"a frequency
$\omega_{bv}$, the third row is the neutrino energy source term $\dot{Q}_\nu$,
and the last row gives the anisotropic velocity $v_a$.
\label{fig:2dmap}}
\end{figure}

Figure \ref{fig:2dmap} shows the evolution of the angle-averaged electron fraction $Y_e$,
Brunt-V\"ais\"al\"a frequency \cite{Buras:2006}, neutrino energy deposition,
and anisotropic velocity \cite{Takiwaki:2012}, giving more insights about the 2D
effects on the different schemes. 
During the early shock expansion (first 100~ms), these profiles are very similar, 
except a slightly noise around the neutrino sphere in the first 50~ms in ASL. 
When convection happens (after $\sim 100$~ms), 
all values close to the shock surface diverge, but the 
central part remains comparable for all schemes and is consistent with the 
previous discussion on Figure~\ref{fig:2D_radial}.

The radial profiles of Brunt-V\"ais\"al\"a frequency and energy deposition 
reveal the different heating behavior between ASL and the other schemes during 
the first $\sim$100~ms. It shows that for ASL there is a lag
of heating between $\sim$50--100~km during the early shock expansion which explains the lower 
maximum shock radius. The lack of this heating feature in ASL is also confirmed by 
the anisotropic velocity profile. 
It shows that convection inside the gain region sets in $\sim$50~ms later in ASL  
than in the other schemes.
Furthermore, the profiles of anisotropic velocity shows that IDSA and ASL tend 
to result in more aspherical flows inside the PNS compared to M1. This is a 
direct consequence of the missing energy transport in the optically thick 
regime typical of leakage schemes, which causes larger entropy gradients in 
spherically symmetric models (Section~\ref{s.1d}), and stronger convection in 
cylindrically symmetric ones. In fact, convection is even stronger in ASL 
compared to IDSA because in the former case leakage prescriptions are assumed 
for all neutrino flavors, while in the latter only for heavy flavor neutrinos. 
This stronger convection  also leads to more noise in the energy source term 
inside the PNS in the ASL simulation: electron 
anti-neutrinos produced in trapped conditions just below the PNS radius are 
advected with the fluid at larger densities, where their presence is suppressed 
by lepton degeneracy. As a consequence, their energy is converted into matter 
internal energy and competes with local neutrino cooling. However, due to the 
high densities inside the PNS, this spurious effect does not translate into 
strong entropy artifacts, but rather in noise in the neutrino energy source 
term, as visible in Figure~\ref{fig:2D_neutrino}. Furthermore, the profile of 
anisotropic velocity reveals that ASL and IDSA evolve shock deformations very 
early. This is visible by the spikes in anisotropic velocity at the shock front 
which is a result of averages considering shocked and unshocked matter. It has 
also been indicated by the radial profiles in Fig.~\ref{fig:2D_radial}, where at 
80~ms (first column) the profile for M1 shows a sharp discontinuity at the shock 
position, but the profiles of ASL and IDSA are slightly smoothed. 
This refers to acoustic waves which evolve from the strong PNS 
convection and aspherical accretion and disturb the spherical symmetry of the 
shock surface in these schemes.

\begin{figure}
\centering
\includegraphics[width=1.0\textwidth]{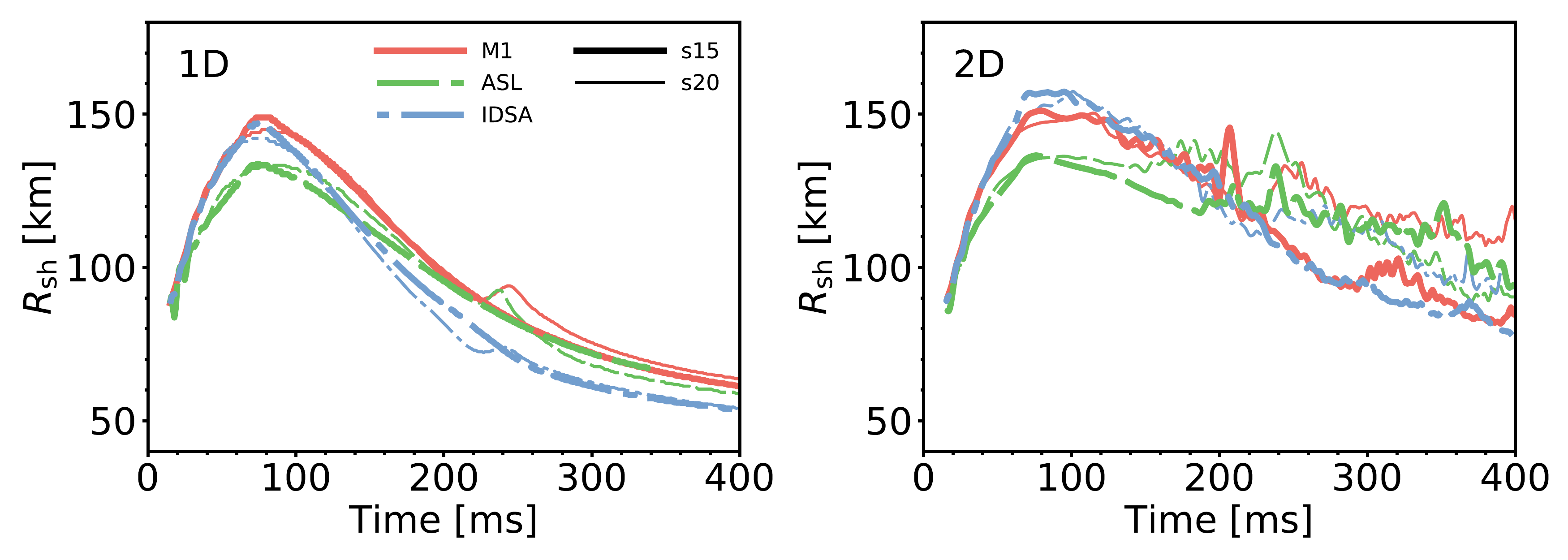}
\caption{Comparison of the average shock radius for a different progenitor
models. The solid lines show the data for the more compact s15 model and the
dashed lines are the already shown results of s20 as a reference. Different
color represents simulations with different neutrino
transport scheme.}
\label{fig:s15}
\end{figure}

A further confirmation of the robustness of this study was done by also varying
the progenitor. Performing the same setup with s15 leads to the same overall
behavior as seen in Figure~\ref{fig:s15}. The s15 progenitor does not have the
accretion of the shell interface at about 220~ms post bounce and the
shock declines much faster which leads to a better overall agreement between
all schemes.

\newpage
%
%
\subsection{Sensitivity study}

Multidimensional simulations using spherical grids \cite{Summa:2016,
Suwa:2013, Takiwaki:2014, Skinner:2016} require small initial density
perturbations to trigger asymmetric motion, e.g.~convection. In contrast, our 2D
simulations have been performed on a cylindrical grid which naturally
introduces perturbations due to the spherical flow on a Cartesian grid. In order to reveal the influence
of these initial perturbations on the simulation outcome, but also the
specific feedback on the different transport schemes, we perform five
additional simulations adding random perturbations at the level of 0.1\% in the
initial post-bounce conditions for each transport scheme. We adopt
the 2D setup with the s20 progenitor and SFHo EoS as described in
Sec.~\ref{sec:2Dcomparison}.

\begin{figure}
 \centering
 \includegraphics[width=0.7\textwidth]{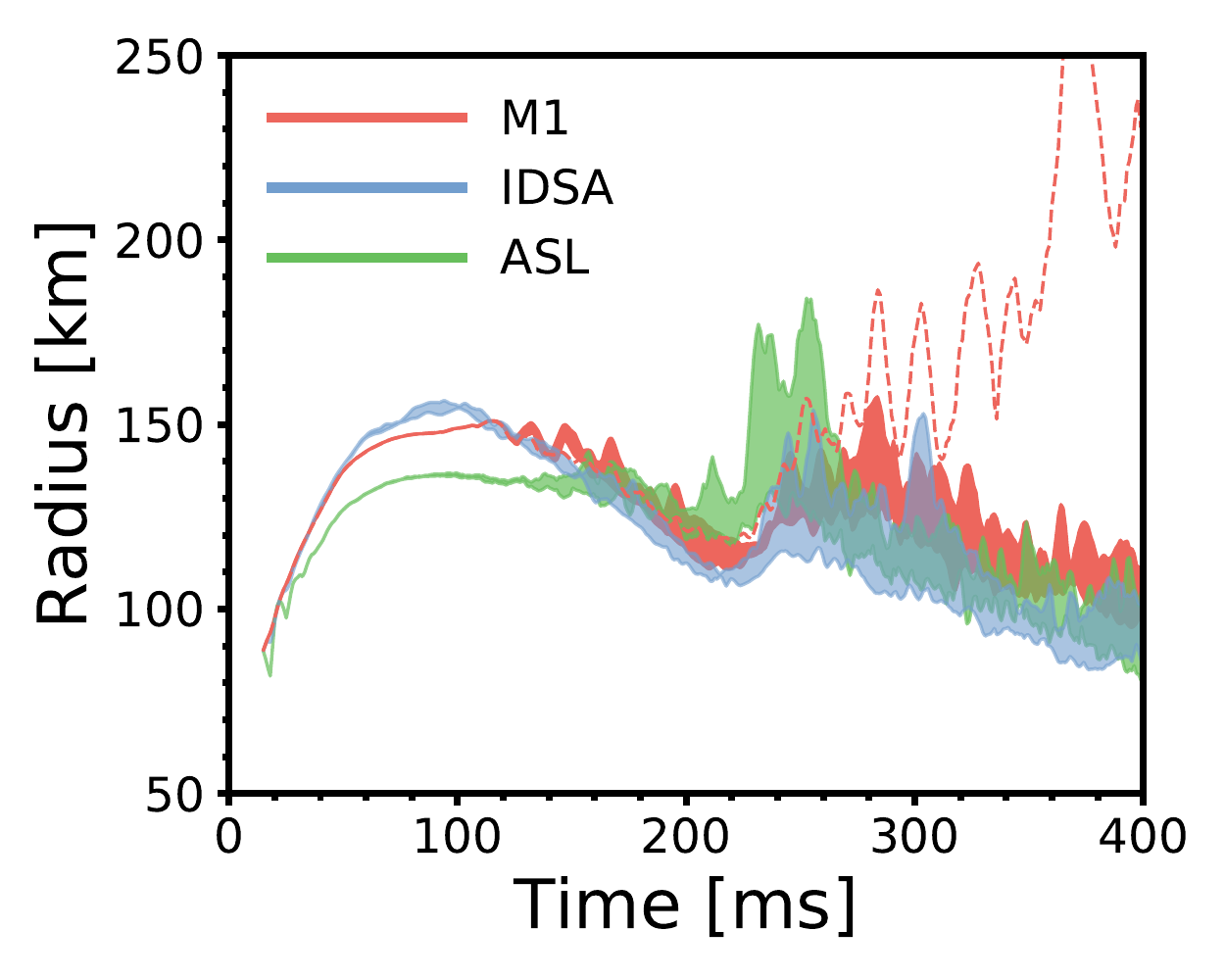}
 \caption{Variation bands of the average shock radius for the 2D simulations
with an introduced initial density perturbation at the per mill level.
Different colors represent different neutrino transport schemes.
The red dashed line shows the only exploding model with M1.
 \label{fig:randomPert}}
\end{figure}

Figure~\ref{fig:randomPert} shows the spread of average shock radii for
simulations with 5 different perturbation seeds for each transport scheme.
It reveals that the influence of these perturbations is very low during the
the shock expansion phase ($\sim 100$~ms). The first visible deviations among
the runs for a given transport scheme (colored bands) are visible for IDSA when
the shock stalls ($\sim 50$~ms). For M1 and ASL the deviations begin growing
later ($\sim 100$~ms). These deviations happened when non-spherical 
transient waves moving around the post-shock region. 
At the moment when the progenitor shell interface crosses the shock ($\sim$220\,ms), 
the sudden re-expansion of the shock further broadens the deviations. 
Even so much as to lead to an explosion in one of the five M1 simulations.  
The large deviations in the runs with ASL at the same time ($\sim$220\,ms) might be
a ray-by-ray effect in 2D which leads to enhanced post-shock fluid
motions and shock deviations as described in \cite{Skinner:2016}, but when the shock declines
again, the deviations shrink and become comparable to the deviations of the
non-exploding bands of the other transport schemes. For these cases, the
averaged trend remains very similar to the results of the first panel in
Fig.~\ref{fig:2D}. As conclusion, we find that IDSA is sensitive to
small perturbations at early times ($\sim$100\,ms), and that the ray-by-ray implementation can 
amplify strong asymmetric shock expansions. 
Inclusion of random perturbations resulted in shock revival and explosion for one M1 simulation that failed otherwise
This points to the overall sensitivity of CCSN simulations to progenitor perturbations, as discussed in detail by
\cite{Couch:2013b,Muller:2015, Muller:2017}. This is especially true in 2D where
stochastic motions can trigger shock expansion which can be very
favourable for the development of an explosion.

%
%
\subsection{2D code performance} \label{sec:performance}
In this section, we give an overview of the computational performance of the different
transport schemes. Our benchmark is designed as follows: we restart the 2D
setup with the s20 progenitor and SFHo EoS as described in
Sec.~\ref{sec:2Dcomparison} at 100~ms post bounce. At this time, the averaged
shock radius is almost at its maximum, see fig.~\ref{fig:2D}, which means that
the initial AMR activity has reached an almost stable configuration.
The observed range spans 100 simulation steps. We compare the ratio of
core-hours spent in the neutrino treatment to the core-hours spent for solving
the hydrodynamics in our simulation. The runs for the different schemes were
performed on different clusters, but this ratio should robustly measure the
performance of the applied scheme. Additionally from the advance in simulation
time during these 100 steps, we extrapolate how many steps the simulation
takes to advance one millisecond in simulation time.

\begin{figure}
 \centering
 \includegraphics[width=0.7\textwidth]{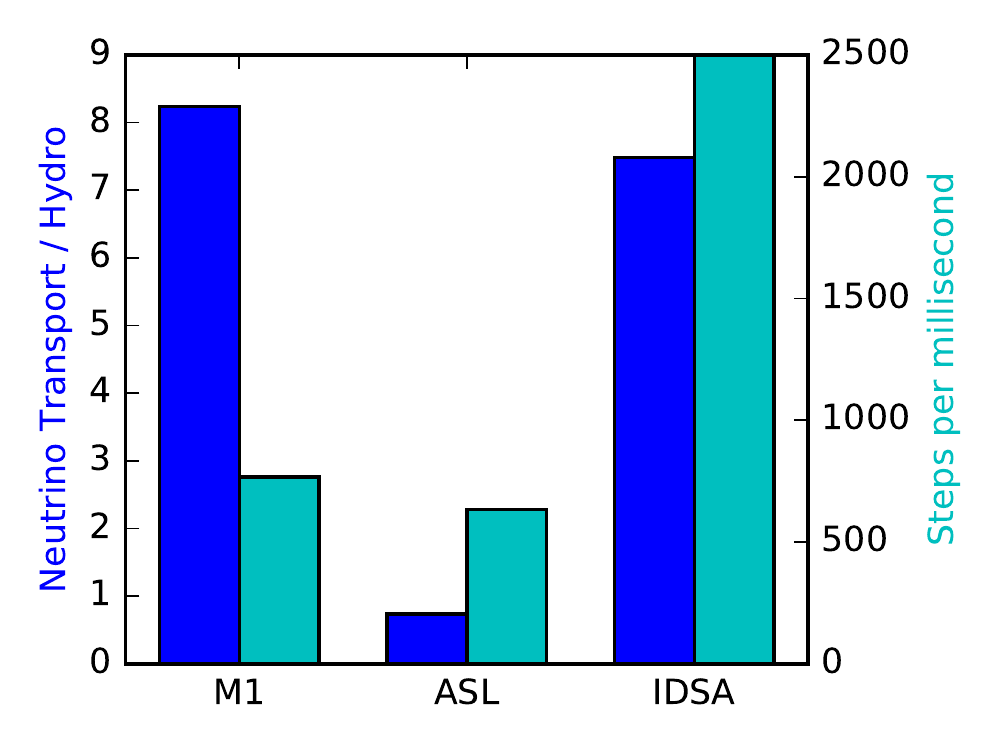}
 \caption{Performance comparison for the different neutrino schemes. The left
axes corresponds to the blue bars (left one for each scheme) showing the ratio
of core-hours spent in Radiation unit compared to the Hydro unit. The right
axis corresponds to the cyan bars giving the number of steps required to
advance the simulation by one millisecond.}
 \label{fig:perf}
\end{figure}

Figure \ref{fig:perf} summarizes the results. As expected, the M1 transport is the most expensive scheme per step, requiring eight times as many core-hours as a single hydrodynamic step.
This is a result of evolving the first two moments
of the neutrino distribution function. M1 is closely followed in expense by IDSA, which requires a factor of
$\sim 7.5$ per hydrodynamic step. 
IDSA spends most of this time in solving the diffusion equation.
ASL is the most approximate scheme, but has the advantage in 
efficiency. When running on a single node, the code spends almost the same
computing time on the hydrodynamic calculation as on the neutrino scheme. 
Regarding the advance in simulation time, M1 and ASL show a comparable
time step restriction which leads to a similar amount of steps to reach 1~ms
simulation time. While the ASL time step is soley based on the CFL
condition for the hydrodynamics (set by the sound speed $c_s$), M1 is set by the CFL condition for
the radiation transport (set by the speed of light $c$).  Nominally, this means a time step difference
of $c/c_s$, but since M1 performs two radiation step per hydrodynamic
step, the ratio is closer to $c/(2c_s)$. Due to the dense regions of
the PNS having a very high sound speed, this ratio is close to unity.
The explicit diffusion solver in IDSA requires a
much smaller time step than M1 and ASL which in the current implementation is
non-adaptive leading to a constant value of 2500 steps per millisecond
simulation time.

The overall 2D performance of the IDSA is worse than M1
but it should be noted that the performance of {\tt FLASH-IDSA} is tuned for 3D
simulations and with GPU acceleration (See~\ref{sec:acc}).
To avoid the overhead of data copying between GPU and CPUs,
we have added an additional layer of AMR block loop by doing data transfer
and neutrino transport at the same time. The sequential calculation on CPUs
leads to the low performance of IDSA in this particular benchmark.

With regard to the memory usage for the different neutrino transport schemes,
IDSA only requires 4 additional variables on the solution per grid cell, 
i.e. 2 ($Y_\nu, Z_\nu$) $\times 2$ ($\nu_e, \bar\nu_e$), 
which refer to the additional conservation equations for trapped neutrino particle and energy fraction, see
Eqs.~(\ref{eq_ynu}) and (\ref{eq_znu}).  
ASL also includes the trapped component of a representative heavy flavor neutrino species
$\nu_x$, which therefore results in $2\times3$ variables per grid cell.
The trapped neutrino spectra in the IDSA and ASL are re-constructed in a AMR block level 
and therefore do not need to be stored in each grid cell. 
On the other hand, M1 carries the spectral neutrino density (scalar) and flux (vector)
on the grid which leads to 4 (density + flux) $\times$ 3 ($\nu_e,
\bar\nu_e, \nu_x$) $\times N_e$ (energy groups), i.e.~in the case of $N_e=12$,
M1 requires 144 additional variables per grid cell. 
In addition to grid variables, IDSA uses $2 (\nu_e, \bar\nu_e) \times N_e (=12) \times R_N (=1000)$ 
variables for the spherically averaged streaming source terms, 
and $9 \times R_N$ spherically averaged thermodynamics variables to solve the streaming component, 
where $R_N$ is the number of radial zones.
ASL uses 37 rays and each ray takes $6 \times R_N (=1000)$ thermodynamics variables. 
These ray-by-ray variables have to be copied and synchronized for each processor 
but it takes much less memory than grid variables.
In 2D simulations, the
memory consumption of the solution usually is not a limiting factor, however it
should be considered in 3D simulations where the number of grid cells
significantly increases and therewith the time and memory consumption for
writing checkpoints.

%
%
\section{Conclusions} \label{sec:conclusions}

We have presented a series of 1D and 2D simulations of the s15 and s20 progenitors
from \cite{Woosley:2007d} with the SFHo, LS220, and HS(DD2) EoS,
and with three different neutrino transport schemes, including M1, IDSA, and ASL.
We ran all these simulations with the publicly available code {\tt FLASH}.
While fixing the hydrodynamics and gravity solvers, we varied the progenitor model,
nuclear EoS, and the neutrino transport scheme in order to investigate 
the impact of different transport methods on features of CCSN simulations.

In spherically symmetric simulations, all three transport schemes show
consistent results on the evolution of the shock and neutrino quantities  
but with variation in certain metrics at about the $\sim$10\% level. 
The variation we observe in 2D simulations is similar to that in
1D, but multidimensional convection leads to larger PNS radii and higher $\mu/\tau$
neutrino luminosities. 
In particular, IDSA and ASL show earlier, stronger PNS convection than M1 
leading to differences in the evolution of the PNS radii and neutrino luminosities.
Between transport schemes, an important difference is the prediction of 
neutrino luminosities and mean energies. 
Especially at later times, these quantities still show a large spread among the schemes.
We find that convection around the PNS surface could produce 
an imbalance of electron and electron anti neutrinos in the ASL 2D run, 
giving large values of electron fraction ($> 0.5$) inside the gain region. 
This could be an artifact from the ray-by-ray implementation in the ASL.

When testing the sensitivity of our results to the initial progenitor profile. 
The differences between the transport schemes show the same trends when varying the progenitor structure and EoS.
When computing resources are limited, our comparison results suggest that
approximate transfer schemes can have value in their potential computational efficiency 
and other key factors such as nuclear EoS, turbulence, dimensionality, etc., 
may result in larger differences than from the neutrino transport approach. 

ASL runs $\sim$10 times faster than M1 and IDSA, making it possible to explore a larger 
parameter space of simulations in 2D (as in \cite{Nakamura:2015})
or even in 3D with reasonably accurate shock dynamics. However, the ASL scheme
seems to inaccurately predict the $Y_e$ evolution, which is sensitive and
important for nucleosynthesis. 
M1 yields the most reliable $Y_e$ evolution and so is preferred for studies concerned with nucleosynthesis.
The IDSA scheme lie between M1 and ASL. The IDSA runs with a slightly
slower speed than M1 but is more memory efficient than M1. 
The memory usage could be a bottleneck for GPU programming and when moving
to 3D simulations. The M1 scheme with three neutrino species accurately
captures the changes of $Y_e$, and the distribution and flux of neutrinos in
both opaque and transparent regions, but at the cost of increased computing time 
and memory.

\ackn
{\tt FLASH} was in part developed
by the DOE NSA-ASC OASCR Flash Center at the University of Chicago.
Analysis and visualization of simulation data were completed using the
analysis toolkit {\tt yt} \cite{Turk:2011}. Partial support for this
work (EO) was provided by NASA through Hubble Fellowship grant
\#51344.001-A awarded by the Space Telescope Science Institute, which
is operated by the Association of Universities for Research in
Astronomy, Inc., for NASA, under contract NAS 5-26555.
S.M.C. is supported by the U.S. Department of Energy, Office of Science, Office of Nuclear Physics,
under Award Numbers DE-SC0015904 and DE-SC0017955 and the Chandra X-ray Observatory under grant TM7-18005X.
A.P. acknowledges support from the INFN initiative ``High Performance data Network'' funded by CIPE.
Helmholtz-University Young Investigator grant No. VH-NG-825, Deutsche
Forschungsgemeinschaft through SFB 1245. Computations
were performed, in part, on resources provided by SNIC (Swedish
National Infrastructure for Computing) through PDC and HPC2N under
Project SNIC 2017/1-259 and, in part, on resources provided by CSCS (Swiss
National Supercomputing Centre) under Project s661, s667 and s840,
and on the high performance computing center (HPCC) at Michigan State University.

\appendix

\section{Relativistic corrections in FLASH-ASL}
\label{Appendix:ASL_relativistic_extension}
The usage of ASL in relativistic simulations or in simulations employing an
effective GR gravitational potential
requires the introduction of the most important relativistic corrections also in the neutrino propagation.
In fact, the ASL results presented in this comparative study are qualitatively
different for pure Newtonian ASL models.
In the latter cases, shock revivals are observed soon after 230ms post bounce,
at the occurrence of the progenitor shell interface, in 2D cylindrically symmetric models.
These explosions are robust with respect to variations in the ASL free parameters
and relate to systematically larger ($\sim$20\%) neutrino mean energies
that significantly enhance the heat deposition inside the gain region.
A similar effect is observed also in 1D, however the increased heating rate is not strong enough to drive an
explosion in more pessimistic spherically symmetric models.

In this appendix, we present the extension of ASL we have adopted in this comparative study, which includes the
gravitational redshift and the Lorentz boost between the fluid and grid
reference frames. They affect radiation propagation in optically thin conditions and its absorption
by the moving fluid. Since in this work we perform 1D spherically symmetric
simulations and, for the 2D cylindrically symmetric
models, we adopt a ray-by-ray approach for the propagation of the free streaming neutrinos, we present
the relativistic extension for spherically symmetric models.

We assume a radial gauge, polar slicing metric
\begin{equation}
 {\rm d}s^2 = - \alpha^2 {\rm d}t^2 +X^{2}{\rm d}R^2 +R^2 \left({\rm d}\theta^2 + \sin^2{\theta}{\rm d}\phi^2  \right),
 \label{eqn: CF metric}
\end{equation}
where $R$ is the areal radius, $\alpha$ the lapse function relating the proper time
lapse of comoving observers to the coordinate time lapse ${\rm d}t$;
$\theta$ and $\phi$ the angles describing a two-sphere; $X = \left( 1 - 2Gm_{\rm grav}/(Rc^2) \right)^{-1/2}$,
$m_{\rm grav}$ being the gravitational mass (i.e., the total energy) enclosed in a sphere of radius $R$.
The lapse function is related with the effective GR gravitational potential $\phi_{\rm GR,eff}$ by:
\begin{equation}
\alpha = \exp{\left( \frac{\phi_{\rm GR,eff}}{c^2} \right)}
\end{equation}
and $\phi_{\rm GR,eff}$ is obtained from the effective gravitational mass $m_{\rm grav}$, as outlined in \cite{OConnor:2015}.

All the local quantities contained inside ASL, including neutrino source terms, are computed in
the fluid reference frame (FRF), distinct from the coordinate frame (CF) associated with the metric Equation~(\ref{eqn: CF metric}).
The neutrino field energy in the two frames are related by a boost
transformation:
\begin{equation}
 \mathcal{E}^{\rm CF} = W (1+v) \mathcal{E}^{\rm FRF},
 \label{eqn: energy boost}
\end{equation}
where $W = (1 - \left( Xv / \alpha  \right)^2)^{1/2}$ is the Lorentz factor, and
$v$ the radial component of the fluid velocity as measured by an observer at constant
radius.
Additionally, the energy of the radiation field, climbing radially out of the gravitational well
by a distance $\Delta R$, is redshifted according to
\begin{equation}
 \frac{\mathcal{E}^{\rm CF}(R)}{\mathcal{E}^{\rm CF}(R + \Delta R)} = \frac{\alpha(R + \Delta R)}{\alpha(R)} \, .
 \label{eqn: gravitational redshift}
\end{equation}
The local spectral neutrino rates $s_{\nu}$ are first transformed from the FRF to the CF.
To design the boost transformation at a coordinate radius $R$ for the spectral rates,
we consider that the amount of emitted neutrinos (per baryonic mass) is Lorentz invariant:
\begin{equation}
 \int_0^{\infty} s_{\nu,{\rm CF}}(R,E) E^2{\rm d}E \, {\rm d}t_{\rm CF}  = \int_0^{\infty} s_{\nu,{\rm FRF}}(R,E) E^2{\rm d}E \, {\rm d}t_{\rm FRF}  \, ,
 \label{eq: boost condition 1}
\end{equation}
where ${\rm d}t_{\rm FRF}$ is the proper time and ${\rm d}t_{\rm FRF} = \alpha~{\rm d}t_{\rm CF}$. The energy in the neutrino field
transforms according to Equation~(\ref{eqn: energy boost}):
\begin{equation}
 \int_0^{\infty} s_{\nu,{\rm CF}}(R,E) E^3{\rm d}E \, {\rm d}t_{\rm CF} = W (1+v)  \int_0^{\infty} s_{\nu,{\rm FRF}}(R,E) E^3{\rm d}E \, {\rm d}t_{\rm FRF}  \, .
 \label{eq: boost condition 2}
\end{equation}
To go from the FRF to the CF, we define $f(R,E) = s_{\nu}(R,E) E^2 $ and make the following ansatz about $f$ in the two frames:
\begin{equation}
 f_{\rm CF}(R,E) = \xi_2  \, f_{\rm FRF}(R,\xi_1 \, E) \, .
 \label{eq: direct boost transformation}
\end{equation}
Then we solve for $\xi_1$ and $\xi_2$ by imposing Equations~(\ref{eq: boost condition 1}) and (\ref{eq: boost condition 2}).

The CF transformed rates are used to evolve radially the neutrino luminosities,
including 
the gravitational redshift. This is done in an operator splitting way: first,
the luminosity is evolved between two neighboring radial zones according to
equation (40) in \cite{Perego:2016}. Then the redshift correction
is applied over the zone separation. We consider that, moving from $R$ to $R + \Delta R$, the particle luminosity
is not affected by the gravitational redshift in the CF frame:
\begin{equation}
 \int_0^{\infty} l_{\nu,{\rm CF}}(R,E) E^2{\rm d}E  = \int_0^{\infty} l_{\nu,{\rm CF}} (R + \Delta R,E) E^2{\rm d}E \, ,
 \label{eq: redshift condition 1}
\end{equation}
while the energy luminosity is (cf.~Equation~(\ref{eqn: gravitational 
redshift})):  

\begin{equation}
 \int_0^{\infty} l_{\nu,{\rm CF}}(R,E) E^3{\rm d}E =   \frac{\alpha(R + \Delta 
R)}{\alpha(R)} \int_0^{\infty} l_{\nu,{\rm CF}}(R + \Delta R,E) 
E^3{\rm d}E  \, .
 \label{eq: redshift condition 2}
\end{equation}
In analogy with the boost transformation, we define $g(R,E) = l_{\nu}(R,E) E^2 $ and we make the following ansatz
about $g$ between $R$ and $R+\Delta R$:
\begin{equation}
 g_{\rm CF}(R+\Delta R,E) = \psi_2 \, g_{\rm CF}(R, \psi_1 \, E) \, .
\end{equation}
We solve for $\psi_1$ and $\psi_2$ by imposing Equations~(\ref{eq: redshift condition 1}) and (\ref{eq: redshift condition 2}).
Finally, the luminosity is locally transformed back to the FRF using the inverse of the boost transformation,
Equation~(\ref{eq: direct boost transformation}), to compute the spectral neutrino
densities required to compute the local absorption rates. This procedure is applied over the entire radial profile.

\section{IDSA performance with GPU acceleration} \label{sec:acc}

\begin{figure}
 \centering
 \includegraphics[width=0.7\textwidth]{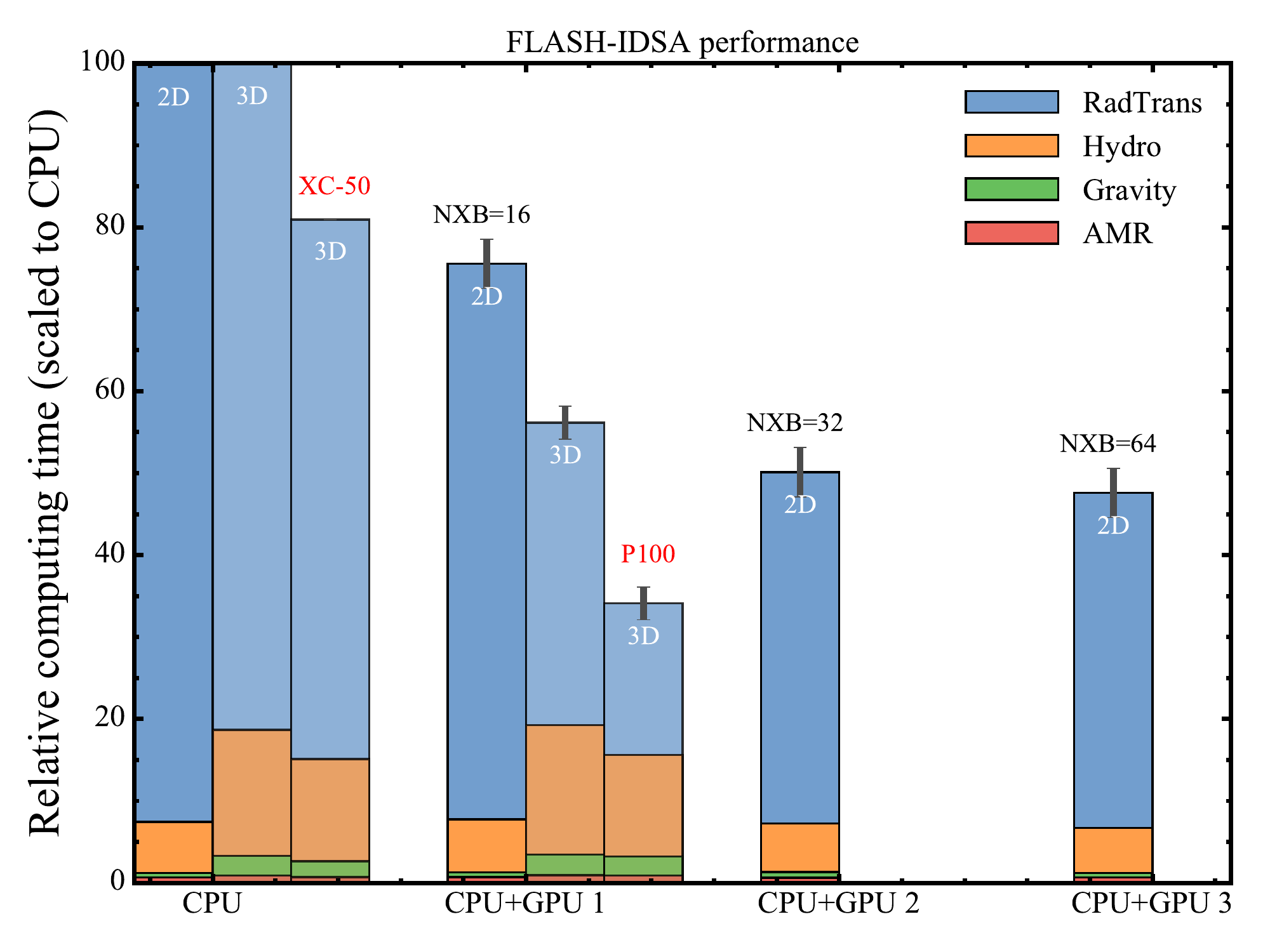}
 \caption{Relative computing time of IDSA on the Piz Daint supercomputer (Cray XC-30 and newly upgraded XC-50).
 Different group of bars represent the performance with/without GPUs and with different AMR block size.
 Bars with(out) red labels are running on the Cray XC-50 (XC-30) system.}
 \label{fig:acc}
\end{figure}

Since the communication in the IDSA diffusion solver is mostly associated with neighboring zones,
we have ported our IDSA solver with {\tt OpenACC} for GPU acceleration.
Figure~\ref{fig:acc} shows the relative computing time of the {\tt FLASH-IDSA}
with different dimension and block size on the Swiss supercomputer, Piz Daint.
The performance is evaluated with 20 energy groups and the baseline run is using
the Cray XC-30 system (with NVIDIA K20X GPU) and with 16 zones per AMR block per dimension.
As discussed in Section~\ref{sec:performance}, the speedup in 2D is worse than 3D.
This is because the 2D data in an AMR block is too small to fill the GPU cores.
Increasing the AMR block size from $16 \times 16$ to $32\times 32$ can further
improve the 2D performance (see Figure~\ref{fig:acc}).
In this study, we need a controlled grid step for all transport schemes to understand the transport effect.
Thus, the grid setup is not tuned to the best 2D performance,
and therefore GPUs are not used in the performance study in Section~\ref{sec:performance}.
The new Cray XC-50 system is about 25\% faster than the original XC-30 system without using GPUs.
The use of P100 GPUs on XC-50 give a speed up of 2.9 in the neutrino transport region and an overall speed up of 2.3.
However even with GPU acceleration, the time step in the current IDSA solver is still restricted to $\sim$4$\times 10^{-7}$\,s
due to the explicit implementation in the diffusion solver.

\section*{References}
\bibliography{flashTransportComp,extraRefs}

\end{document}